\documentclass[onecolumn,a4paper,published,11pt,accepted=2024-06-26]{quantumarticle}
\pdfoutput=1

\usepackage[utf8]{inputenc}
\usepackage{graphicx}
\usepackage{hyperref}
\usepackage{amsmath}
\usepackage{amssymb}
\usepackage{xcolor}
\usepackage{amsthm}
\usepackage{subcaption}
\usepackage{cleveref}
\usepackage{tikz}
\usepackage{quantikz}
\usepackage[numbers,sort&compress]{natbib}

\title{Error-corrected Hadamard gate simulated at the circuit level}

\author{Gy\"{o}rgy P. Geh\'{e}r} 
\email{gehergyuri@gmail.com, george.geher@riverlane.com}
\affiliation{Riverlane, Cambridge, CB2 3BZ, UK}

\author{Campbell McLauchlan}
\email{campbell.mclauchlan@gmail.com}
\affiliation{Riverlane, Cambridge, CB2 3BZ, UK}
\affiliation{DAMTP, University of Cambridge, Wilberforce Road, Cambridge, CB3 0WA, UK}

\author{Earl T. Campbell}
\email{earl.campbell@riverlane.com}
\affiliation{Riverlane, Cambridge, CB2 3BZ, UK}
\affiliation{Department of Physics and Astronomy, University of Sheffield, Sheffield S3 7RH, UK}

\author{Alexandra E. Moylett}
\email{alex.moylett@riverlane.com}
\affiliation{Riverlane, Cambridge, CB2 3BZ, UK}

\author{Ophelia Crawford}
\email{ophelia.crawford@riverlane.com}
\affiliation{Riverlane, Cambridge, CB2 3BZ, UK}

\begin{document}

\maketitle

\begin{abstract}
    We simulate the logical Hadamard gate in the surface code under a circuit-level noise model, compiling it to a physical circuit on square-grid connectivity hardware. Our paper is the first to do this for a logical unitary gate on a quantum error-correction code. We consider two proposals, both via patch-deformation: one that applies a transversal Hadamard gate (i.e. a domain wall through time) to interchange the logical $X$ and $Z$ strings, and another that applies a domain wall through space to achieve this interchange. We explain in detail why they perform the logical Hadamard gate by tracking how the stabilisers and the logical operators are transformed in each quantum error-correction round. We optimise the physical circuits and evaluate their logical failure probabilities, which we find to be comparable to those of a quantum memory experiment for the same number of quantum error-correction rounds. We present syndrome-extraction circuits that maintain the same effective distance under circuit-level noise as under phenomenological noise. We also explain how a $SWAP$-quantum-error-correction round (required to return the patch to its initial position) can be compiled to only four two-qubit gate layers. This can be applied to more general scenarios and, as a byproduct, explains from first principles how the ``stepping'' circuits of the recent Google paper \cite{McEwenBaconGidney} can be constructed. 
\end{abstract}

\section{Introduction}

Execution of useful quantum algorithms with reliable results requires quantum error correction (QEC) to correct for physical noise during the computation. QEC works by measuring a set of stabilisers (Pauli observables) repeatedly \cite{Gottesman-thesis,terhal2015quantum,campbell2017roads,roffe2019quantum} and, based on the measurement outcomes, applying certain corrections either directly to the physical qubits or in software. As a first step to examining the performance of any QEC code, it is natural to evaluate how well encoded quantum states can be stored through time -- a procedure called quantum memory. This can be implemented efficiently on a classical computer, using a Clifford simulator \cite{Gottesman-thesis,stim}, where we may choose between two main categories of noise models: a phenomenological Pauli noise model, and a circuit-level Pauli noise model. Phenomenological noise makes the unrealistic assumption that the stabilisers can be measured directly. Meanwhile, circuit-level noise assumes stabilisers are measured using syndrome extraction circuits composed of single-qubit measurements, auxiliary qubit state preparation, and one- and two-qubit gates, with noise on each component operation. Such syndrome extraction circuits must respect the connectivity constraints of the hardware implementing the experiment. Performance of QEC codes varies dramatically depending on the chosen noise model. For instance, the rotated planar surface code, which is among the leading QEC code candidates, has a threshold of $\sim 2.9\%$ under phenomenological noise, and of $\sim 0.8\%$ under circuit-level noise~\cite{Raussendorf_2007,fowler2012surface}.

To estimate the performance of QEC codes during fault-tolerant quantum computation (FTQC), we also need to evaluate the performance of logical gates. While quantum memory experiments have been simulated many times in the literature for various codes under circuit-level noise~\cite{Raussendorf_2007,fowler2012surface,Tomita_2014,chamberland2020topological,higgott2023improved}, logical gate simulations are not as comprehensive. Two-qubit Pauli measurements realised by lattice surgery have been simulated at the circuit level, both with twists \cite{ChamCamtwistbased} and without \cite{ChamCamtwistfree}. However, code performance under logical Hadamard (and other unitary) operations remains relatively unexplored. This is a reasonable next step since, despite only being a single logical qubit operation, the logical Hadamard gate involves more steps than a lattice surgery operation (see \Cref{sec:explanatation} and Ref.~\cite{Horsman_2012, ChamCamtwistfree, ChamCamtwistbased}).

Horsman \textit{et al.}  \cite{Horsman_2012} discussed the logical Hadamard gate for the unrotated planar surface code and highlighted that it is not sufficient to perform a transversal Hadamard gate; we quote, ``Performing a Hadamard gate transversally by Hadamard operations on each individual qubit will leave us with a planar qubit that is in the correct state \dots, but it will leave the planar surface at a different orientation from the original'' \cite[Section 4.3]{Horsman_2012}. This issue persists for the lately more popular rotated planar code case, as a transversal Hadamard changes the so-called background lattice (see e.g. \cite[Appendix A.1]{tangled_schedules}), which in turn changes the type of lattice surgery operations that can be executed with low-depth circuits while respecting device connectivity. Therefore, it is desirable to perform the logical Hadamard gate such that it leaves the code patch in the exact position in which it started, but with interchanged $X$ and $Z$ logical operators. 

We also point out that, even though the popular Pauli-based computational (PBC) model \cite{BravyiPBC} for FTQC uses only a series of multi-qubit Pauli measurements without the need to implement any logical unitary gates, it does impose constraints on hardware connectivity and the construction of effective syndrome extraction circuits~\cite{ChamCamtwistfree,ChamCamtwistbased,tangled_schedules}. Furthermore, these measurements must typically be performed sequentially, meaning circuit run-time is directly proportional to the number of non-Clifford gates present. On the other hand, the alternative computational model, where we execute each logical gate, can be implemented straightforwardly on the surface code under the natural square-grid connectivity \cite[Sec. III.]{blunt2023compilation} and possibly enables greater parallelism \cite{gidney2021factor,litinski2022active} as the gates are typically restricted to a small number of qubits. However, we note that, in general, the tradeoffs are not entirely understood.

We consider two proposals to execute the logical Hadamard gate.  The first construction is similar to that of Bombin \textit{et al.} \cite{Bombin2021} and is composed of transversal Hadamard gates, a series of patch-deformations, and then two layers of $SWAP$ gates to return the patch to its original position.  Their simulations consider phenomenological noise, whereas we take a circuit-level approach. The second construction is a new one that also applies patch-deformation, but using a domain wall instead of transversal Hadamards. The domain wall introduces stabilisers of mixed $X$-$Z$ type, with which we achieve the interchanging of the logical operators.  Indeed, both approaches can be thought of as exploiting domain walls if we think of transversal Hadamards as forming a ``time-like'' domain wall, and thus they are topologically equivalent.

In this paper, we simulate the logical Hadamard gate implemented on the rotated planar code on a square-grid layout hardware, for the first time (to the best of our knowledge) accounting for full circuit-level noise. Crucially, the circuit-level paradigm introduces various subtleties that are absent in the phenomenological setting. For instance, we must carefully construct the syndrome extraction circuit so that it does not reduce the effective (or fault) distance, i.e. it does not decrease the minimum number of fault locations necessary to cause an undetectable logical failure. The number of necessary QEC rounds may also have to be increased slightly, as under circuit-level noise we introduce so-called hook errors that have both space and time components.

The outline of the paper is as follows. In the next section, we explain both methods for the logical Hadamard gate procedure in detail, but without taking into account the circuit itself. In particular, we explain how the logical operators are transformed in each step. \Cref{sec:scheduling} explains the syndrome extraction circuits and the number of QEC rounds that are chosen for each step to maintain the full effective distance $d$ against all types of mid-circuit faults. The logical Hadamard experiment involves two $SWAP$-QEC rounds to shift the patch in a diagonal direction by first swapping qubits and then measuring the stabilisers of the shifted patch. These $SWAP$-QEC rounds are optimised in \Cref{sec:compilation} to use only four layers of two-qubit gates, like a standard QEC round. This optimisation provides an alternative derivation of the recently-proposed ``stepping'' circuits \cite{McEwenBaconGidney}. In \Cref{sec:simulation}, we numerically evaluate the QEC performance of the logical Hadamard experiment and compare it to the standard quantum memory experiment performed on the original patch with the same number of QEC rounds. While the Hadamard circuit is larger, with more possible fault locations, it can tolerate some higher-weight errors. We find these effects balance out, so the Hadamard and memory experiments have similar performance, with the slight gap closing as the code distance increases or the physical noise decreases. We conclude the main part of the paper with some discussion in \Cref{sec:conclusion}. The appendices contain further supplementary material.


\section{Explanation of the two methods}\label{sec:explanatation}

\begin{figure}
    \centering
    \includegraphics[width=0.7\textwidth]{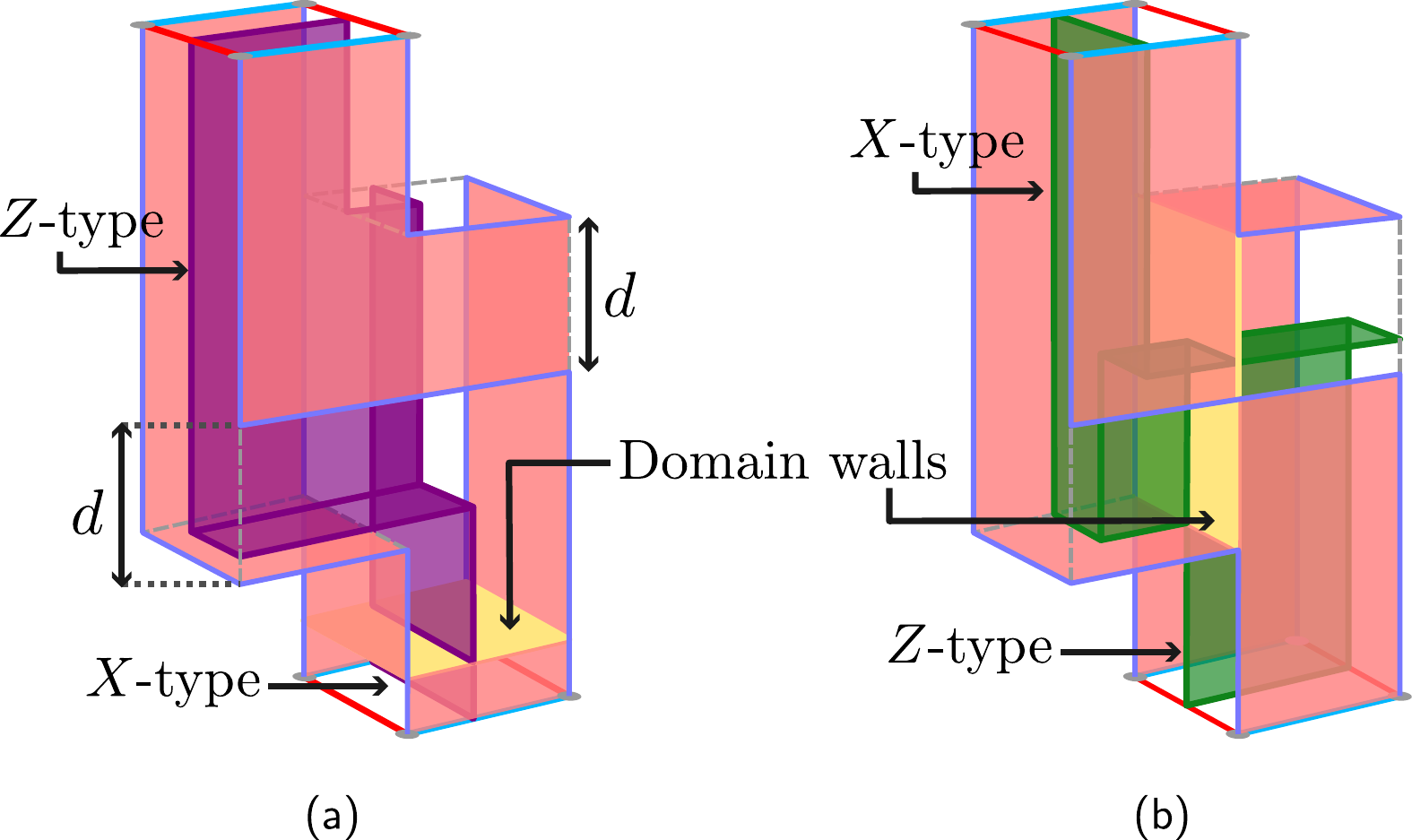}
    \caption{Schematic illustration of the procedures introduced in this section via space-time diagrams (time running upwards), up until the movement of the patches to their original locations (see \Cref{fig:log_had_transversal_patches} and \Cref{fig:log_had_dom_wall_patches}). In (a), we illustrate the transversal Hadamard procedure, where the transversal Hadamard is indicated by a yellow, time-like domain wall. We display the evolution of one of the logical observables, coloured purple, which transforms from $X$-type to $Z$-type. Boundaries of the patch are surfaces coloured red if $X$-type operators can terminate there, or left clear otherwise. The patch deformation steps are shown to be separated in time by $d$ QEC rounds. In (b), we illustrate the domain wall patch deformation procedure, with the (now space-like) domain wall again indicated by a yellow surface. Here, we display the evolution of the other logical observable, coloured green, which transforms from $Z$-type to $X$-type.}
    \label{fig:ST_diagrams}
\end{figure}

In this section, we detail two methods for implementing a logical Hadamard gate in a square surface code patch. In both cases, we demonstrate how the logical $X$ and $Z$ operators are interchanged, achieving both an exchange of positions and also an exchange of $X$ and $Z$ strings. While the positions of the logical strings can be interchanged simply via patch deformation~\cite{GoSc}, this leaves the Pauli types of the logical strings unchanged. To affect the required change, we must insert a domain wall into the patch, which has the effect of exchanging the Pauli bases (as $X\leftrightarrow Z$) of any string operators that cross it. The two methods outlined below differ in the type of domain wall that is inserted. In \Cref{sec:Log_Had_Transversal}, we describe a method illustrated in \Cref{fig:log_had_transversal_patches}, which involves applying a transversal Hadamard gate to all qubits. This is equivalent to inserting a ``time-like" domain wall at some time across the entire patch; see also e.g. \cite{Bombin2021}. In \Cref{sec:Log_Had_DWPD}, we detail a method in which a domain wall is inserted along a line in the patch, illustrated in \Cref{fig:log_had_dom_wall_patches}. Both methods can be seen to be closely related (in \Cref{sec:simulation} we show that compiling to a particular choice of native gates can result in equivalent circuits for the two procedures) and, indeed, both critically involve the braiding of the code's logical corners via patch deformation, as is explained in \Cref{app:Corner_Braiding}. In \Cref{fig:ST_diagrams} we present schematic space-time illustrations to summarise the two procedures. 

\subsection{Logical Hadamard gate via transversal Hadamard and patch deformation}\label{sec:Log_Had_Transversal}

\begin{figure}
     \centering
     \begin{subfigure}[b]{0.3\textwidth}
         \centering
         \includegraphics[width=0.75\textwidth]{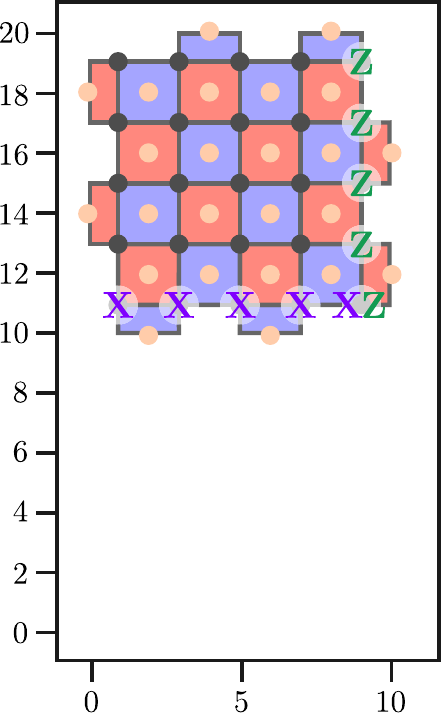}
         \caption{}
         \label{fig:LHTrans_a}
     \end{subfigure}
     \hfill
     \begin{subfigure}[b]{0.3\textwidth}
         \centering
         \includegraphics[width=0.75\textwidth]{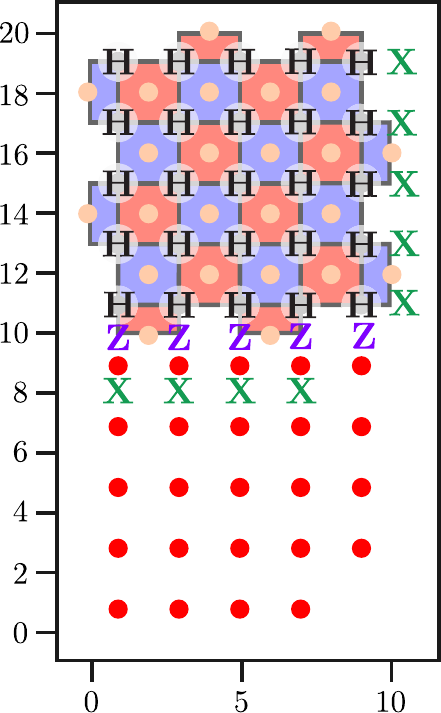}
         \caption{}
         \label{fig:LHTrans_b}
     \end{subfigure}
     \hfill
     \begin{subfigure}[b]{0.3\textwidth}
         \centering
         \includegraphics[width=0.75\textwidth]{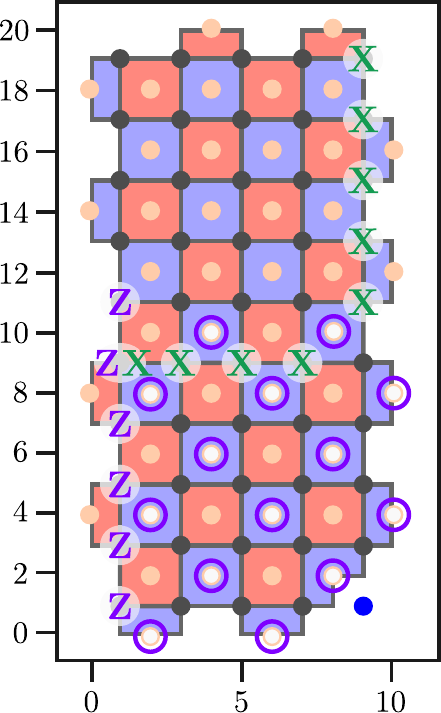}
         \caption{}
         \label{fig:LHTrans_c}
     \end{subfigure}
     \hfill
     \begin{subfigure}[b]{0.3\textwidth}
         \centering
         \includegraphics[width=0.75\textwidth]{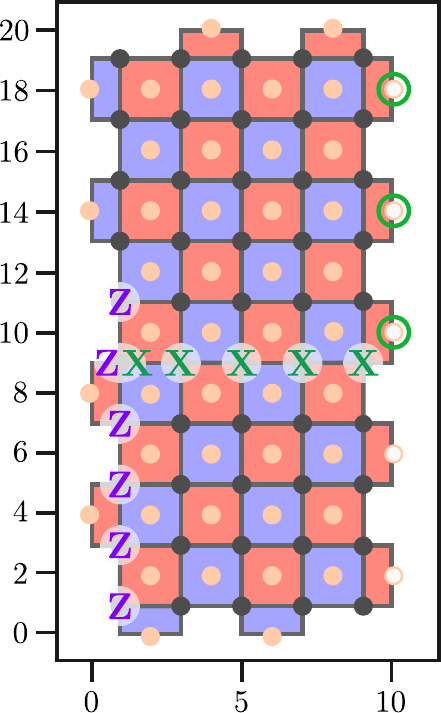}
         \caption{}
         \label{fig:LHTrans_d}
     \end{subfigure}
     \hfill
     \begin{subfigure}[b]{0.3\textwidth}
         \centering
         \includegraphics[width=0.75\textwidth]{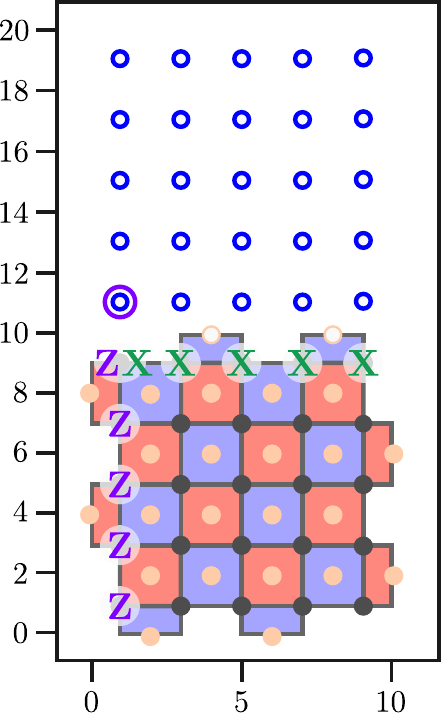}
         \caption{}
         \label{fig:LHTrans_e}
     \end{subfigure}\hfill
     \begin{subfigure}[b]{0.3\textwidth}
         \centering
         \includegraphics[width=0.75\textwidth]{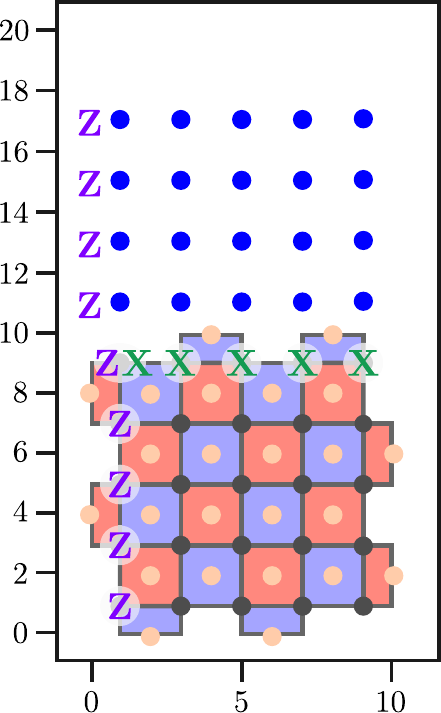}
         \caption{}
         \label{fig:LHDW_f}
     \end{subfigure}
     \hfill
     \begin{subfigure}[b]{0.3\textwidth}
         \centering
         \includegraphics[width=0.75\textwidth]{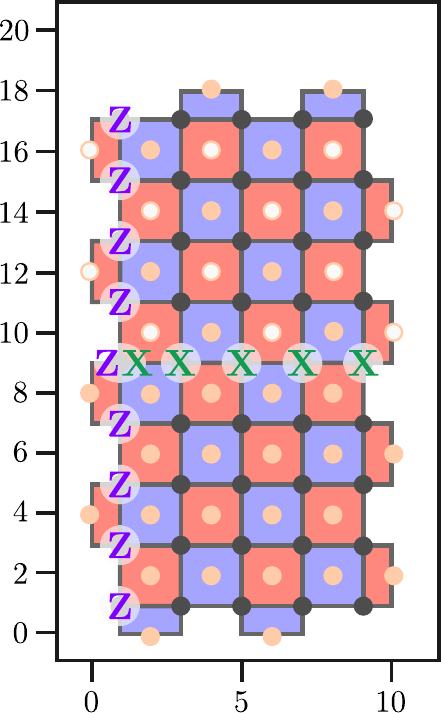}
         \caption{}
         \label{fig:LHDW_g}
     \end{subfigure}
     \hfill
     \begin{subfigure}[b]{0.3\textwidth}
         \centering
         \includegraphics[width=0.75\textwidth]{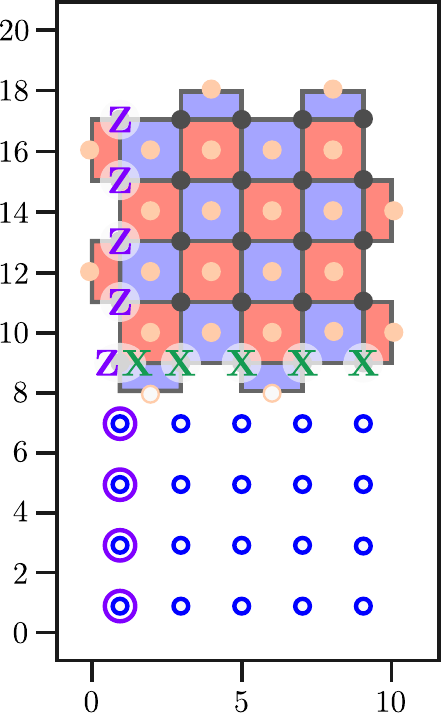}
         \caption{}
         \label{fig:LHDW_h}
     \end{subfigure}
     \hfill
     \begin{subfigure}[b]{0.3\textwidth}
         \centering
         \includegraphics[width=0.75\textwidth]{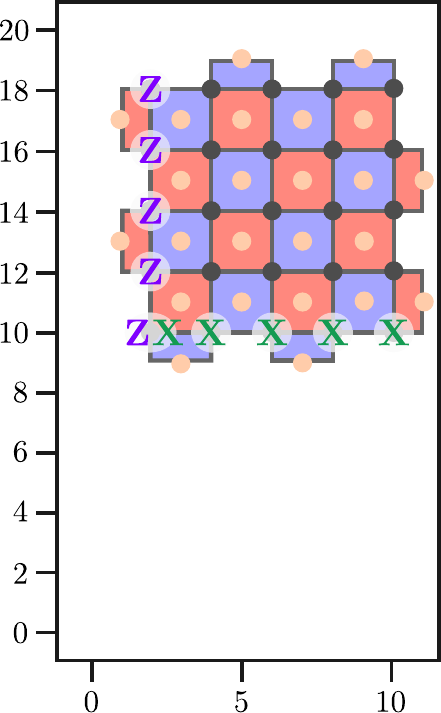}
         \caption{}
         \label{fig:LHDW_i}
     \end{subfigure}
     \hfill
        \caption{Logical Hadamard implementation steps for the method in which a transversal Hadamard is applied to all data qubits (sub-figure (b)). Data qubits (auxiliary qubits) are indicated in grey (orange). Blue (red) plaquettes represent $Z$ ($X$)-type stabilisers. Each white circle represents a non-deterministic measurement outcome the first time the indicated stabiliser/qubit is measured. Qubits ringed in blue (red) are measured in the $Z$ ($X$) basis in that step. Solid blue (red) qubits are initialised in the $Z$ ($X$) basis. We keep track of the evolution of logical operators by colouring them purple or green respectively. Purple (green) outer rings around a measured data/auxiliary qubit indicate measurement outcomes that are multiplied into the purple (green) logical operator resulting in the updated operator.}
        \label{fig:log_had_transversal_patches}
\end{figure}

We illustrate in \Cref{fig:log_had_transversal_patches} how the logical Hadamard gate is implemented using a transversal Hadamard gate as a ``time-like" domain wall in the patch. This is the more common route found in the literature for performing the logical Hadamard gate (see e.g. \cite{Bombin2021}). Throughout the paper, stabiliser plaquettes are coloured red if they are $X$-type stabilisers, blue if they are $Z$-type stabilisers, and half-blue, half-red if they are mixed, with $Z$ operators located in the blue part of the plaquette and $X$ operators in the red part. The purple string operators in each figure show the current state of the $X$-logical operator of the initial patch, and the green strings show the same for the $Z$-logical operator. We explain each step below.

We begin with a square patch of $d\times d$ surface code encoding a single logical qubit (\Cref{fig:LHTrans_a} shows a $d=5$ example; the method is the same for any $d\times d$ patch where $d\geq 2$ is arbitrary). We could perform one or more QEC rounds of syndrome extraction on this patch, but these will not be counted as part of the logical Hadamard procedure. We then apply a Hadamard gate to all data qubits and prepare a further array of $d^2 - 1$ qubits below the initial patch in the $|+\rangle$ state (\Cref{fig:LHTrans_b}). Notice that the transversal Hadamard exchanges the roles of all $X$/$Z$ plaquettes and changes the bases of the two logical strings. For all subsequent steps, as is usual in patch deformation, the logical information is preserved if we can find representatives of each logical operator that commute with the measurements performed. Each panel in \Cref{fig:log_had_transversal_patches} shows those logical strings that commute with that round of measurements, along with the measurement outcomes (circled in purple/green) that were multiplied into the logical string to deform it to its new position. 

After performing the transversal Hadamard gate, we extend the patch vertically to a $2d\times d$ sized patch, as shown in \Cref{fig:LHTrans_c}. This is done by measuring the stabilisers of the new patch for $d$ QEC rounds (see \Cref{sec:distance_repeat_mmts} for an explanation of the number of QEC rounds at each step). White circles in the figure indicate measurements with non-deterministic outcomes. On the one hand, the $Z$ string from \Cref{fig:LHTrans_b} commutes with the extra stabiliser measurements. Those whose auxiliary qubits are circled in purple are used to deform this $Z$ string to the representative shown in \Cref{fig:LHTrans_c}. On the other hand, some of the new stabiliser measurements do not commute with the green logical string. However, note that the equivalent, green, extended $X$ string shown in \Cref{fig:LHTrans_b} and \ref{fig:LHTrans_c} does commute with the new stabilisers for the patch. Since the added part of the $X$ string stabilised the state of the system before the stabiliser measurements of the extended patch (owing to the state preparation chosen), the extended string is a representative of the same logical operator as the original string. Thus, the information is preserved by the stabiliser measurements.

We then change the stabilisers along the right boundary of the patch (\Cref{fig:LHTrans_d}), performing another $d$ rounds of syndrome extraction. The green logical string is transformed by multiplying in the outcomes of those new stabiliser measurements in the top half of the patch (circled in green in \Cref{fig:LHTrans_d}). We then shrink the patch from the top to return it to its original square shape. We do this by measuring all $d^2$ qubits in the top half of the patch in the $Z$ basis (\Cref{fig:LHTrans_e}). We use the results of these single-qubit measurements and the previous results of $Z$ stabiliser measurements to form detectors (see \Cref{sec:scheduling}). Simultaneously, we perform one further round of syndrome extraction for the remaining stabilisers. The purple logical operator can be shrunk by one qubit (circled in purple) as we measure qubit $(1,11)$ in the $Z$ basis. As a result of these measurements, the green operator has moved from the right-hand boundary of the patch to the top boundary, while the purple operator has moved from bottom boundary to left-hand boundary, and the Pauli bases of these two operators have swapped. We have nearly implemented a logical Hadamard gate, except the patch is not in its original location.

To move the patch to its original location, we begin by initialising $d(d-1)$ of the previously measured qubits in the $|0\rangle$ state, as shown in \Cref{fig:LHDW_f}. We then perform patch extension: we measure a set of stabilisers that extends the patch upwards to form a $(2d-1)\times d$ sized patch (\Cref{fig:LHDW_g}) and perform $d$ rounds of syndrome extraction for this new stabiliser set. The logical operator representatives that commute with these patch extension measurements are shown in \Cref{fig:LHDW_g}. Finally, we perform patch shrinking again, measuring the $d(d-1)$ qubits in the bottom half of the patch in the $Z$ basis (\Cref{fig:LHDW_h}), forming detectors from these measurements, and simultaneously perform one QEC round of syndrome extraction for the top half. This has moved the patch to a position one data qubit below its original location.

Before we rectify this slight displacement, notice that we cannot simply extend/shrink the patch to move it one data qubit higher. Indeed, the resulting patch would not be that of \Cref{fig:LHTrans_a}. Instead, its weight-two stabilisers would be shifted and the positions of bulk $X$ and $Z$ plaquettes would be interchanged. This is a result of having inserted the time-like domain wall into the patch at \Cref{fig:LHTrans_b}. To circumvent this, we instead apply a circuit that shifts the patch one data qubit higher. Namely, we apply a $SWAP$ gate between each qubit at location $(x,y)$ with that at location $(x+1, y+1)$, interchanging the roles of data and auxiliary qubits. We then perform one QEC round of syndrome extraction, using the stabilisers shown in \Cref{fig:LHDW_i}. Finally, we move the patch again by performing a $SWAP$ gate between qubits $(x,y)$ and $(x-1,y+1)$, once again interchanging the roles of data and auxiliary qubits. This patch is now the same as the original patch from \Cref{fig:LHTrans_a}, but with a logical Hadamard gate having been implemented. We may perform one QEC round to finish the logical Hadamard experiment.

\subsection{Logical Hadamard gate via domain wall patch deformation}\label{sec:Log_Had_DWPD}

\begin{figure}
     \centering
     \begin{subfigure}[b]{0.18\textwidth}
         \centering
         \includegraphics[width=\textwidth]{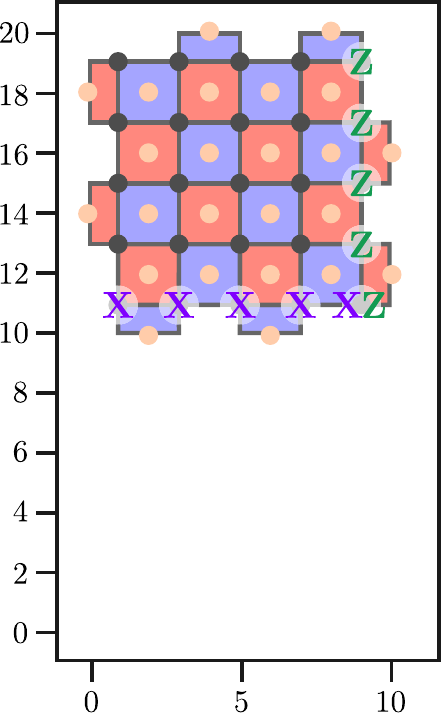}
         \caption{}
         \label{fig:LHDW_a}
     \end{subfigure}
     \hfill
     \begin{subfigure}[b]{0.18\textwidth}
         \centering
         \includegraphics[width=\textwidth]{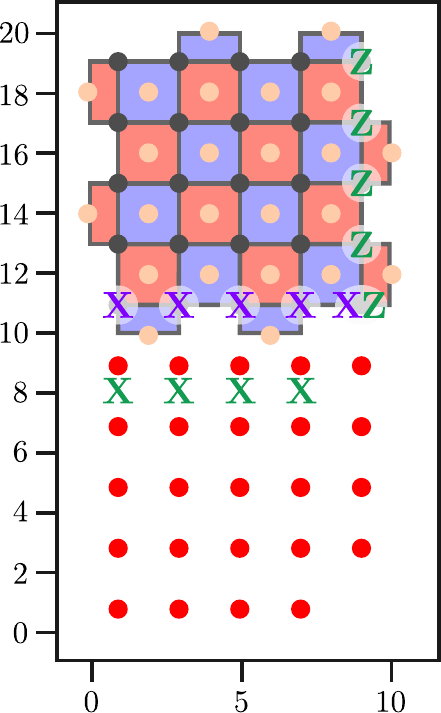}
         \caption{}
         \label{fig:LHDW_b}
     \end{subfigure}
     \hfill
     \begin{subfigure}[b]{0.18\textwidth}
         \centering
         \includegraphics[width=\textwidth]{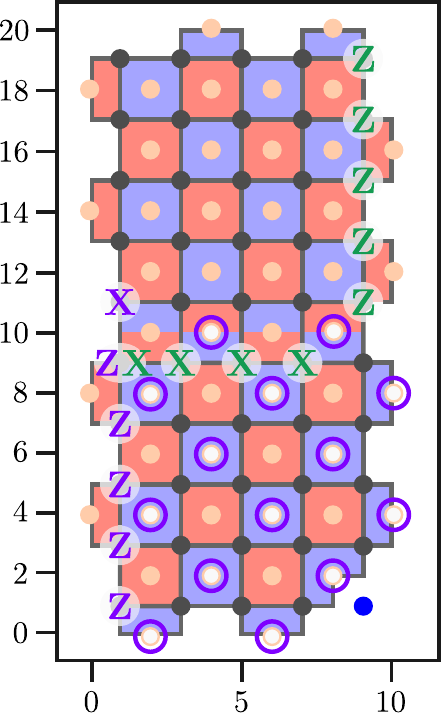}
         \caption{}
         \label{fig:LHDW_c}
     \end{subfigure}
     \hfill
     \begin{subfigure}[b]{0.18\textwidth}
         \centering
         \includegraphics[width=\textwidth]{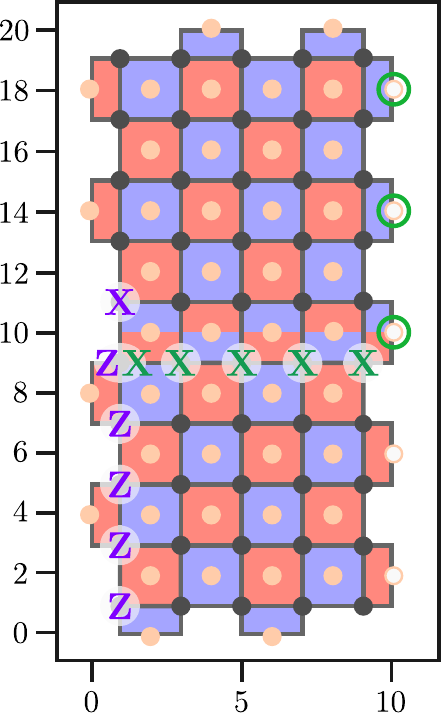}
         \caption{}
         \label{fig:LHDW_d}
     \end{subfigure}
     \hfill
     \begin{subfigure}[b]{0.18\textwidth}
         \centering
         \includegraphics[width=\textwidth]{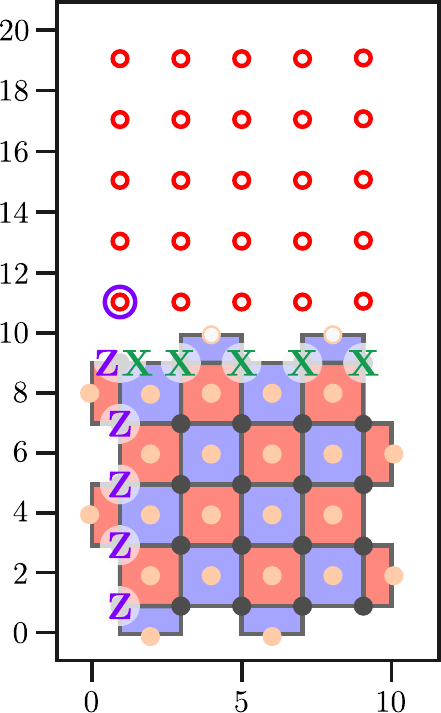}
         \caption{}
         \label{fig:LHDW_e}
     \end{subfigure}
     \hfill
        \caption{Logical Hadamard gate method in which a domain wall is inserted into the patch (sub-figures (c) and (d)). Mixed blue-red plaquettes act as $Z$ ($X$) on the qubits intersecting the blue (red) part of the plaquette. The steps after sub-figure (e) are analogous to those in \Cref{fig:LHDW_f}-\ref{fig:LHDW_i}.}
        \label{fig:log_had_dom_wall_patches}
\end{figure}

We now explain the alternative method that utilises a ``space-like'' domain wall running along the width of the patch, in place of the transversal Hadamard gate of the previous proposal. We begin with the same patch as before (\Cref{fig:LHDW_a}). We again initialise $d^2 - 1$ additional data qubits below the initial patch in the $X$ basis. However, this time we do not apply any Hadamard gates (\Cref{fig:LHDW_b}).

We continue by performing patch extension, but now with a domain wall in the resulting patch, as shown in \Cref{fig:LHDW_c}. The domain wall is located at the line $y=10$. Notice that any logical strings that cross this domain wall have mixed $X$/$Z$ type. The measurements of \Cref{fig:LHDW_c} have not only resulted in this domain wall being inserted into the patch, but have also moved the logical corners (locations where the boundary type changes). In particular, the corners are now located at positions $(1,1)$, $(1,11)$, $(1,19)$ and $(9, 19)$. This, indeed, occurs in the previous proposal as well. In \Cref{app:Corner_Braiding}, we explain how the movements of the corners realize a ``braid'' that enacts the logical Hadamard gate. Similarly to the previous proposal, we also transform the logical operators to those shown in \Cref{fig:LHDW_c}, performing $d$ QEC rounds of syndrome extraction at this point. 

Again, we measure new stabilisers on the right-hand boundary of the patch (\Cref{fig:LHDW_d}) for $d$ QEC rounds, updating the green logical string to that shown in the figure. Finally, we perform patch shrinking by measuring the $d\times d$ qubits in the top half of the patch, this time in the $X$ basis. The purple logical operator has its weight reduced by one, leaving the patch and logical operators shown in \Cref{fig:LHDW_e}. As can be seen from the figure, these measurements result in pure $X$- and $Z$-type logical operators. Note that the reason for the different measurements in \Cref{fig:LHDW_e}, compared to the previous method (\Cref{fig:LHTrans_e}), is the fact that we do not apply a transversal Hadamard, which resulted in an interchange of $X$ and $Z$ in all subsequent measurements and logical operators in this top half of the patch. After the single-qubit measurements, the logical Hadamard gate has been implemented, except the patch is displaced, as in the previous method. This then is followed by the same procedure as in the previous proposal (starting with \Cref{fig:LHDW_f}) for moving the patch back to its original location.


\section{Maintaining code distance under circuit-level noise}\label{sec:scheduling}

When implementing the logical Hadamard gate experiment on hardware, we need to construct the circuits that extract the syndromes of stabilisers. During this construction, it is crucial that each stabiliser is measured with a circuit that maintains the code's effective distance $d$. To achieve this, we need to select an appropriate schedule \cite{tangled_schedules} for the entangling gates used to perform stabiliser measurements. Furthermore, we must ensure that the distance under measurement faults remains $d$ as well. Below, we begin by explaining why, to maintain distance $d$ under faulty measurements, the above two proposals involve steps that include $d$ QEC rounds. We will then describe the chosen scheduling of the entangling gates in our syndrome extraction circuits that maintains the effective distance under circuit-level Pauli noise.

\subsection{Maintaining distance through repeated measurements}\label{sec:distance_repeat_mmts}

During the execution of the logical Hadamard gate via both proposed methods, there are stages at which we extend the surface code, or move the logical corners. These steps involve stabiliser measurements with non-deterministic outcomes (white circles in \Cref{fig:log_had_transversal_patches} and \ref{fig:log_had_dom_wall_patches}). In some of these cases, $d$ rounds of syndrome extraction are required to maintain the effective code distance. 

The first round of syndrome extraction for the logical Hadamard procedure occurs during patch extension (\Cref{fig:LHTrans_c} and \Cref{fig:LHDW_c}). Because the purple logical string is transformed after this process, we must measure $d$ times all stabilisers that are multiplied into the string to form the new logical representative. This is because, if any of these stabiliser measurements are recorded incorrectly during all $d$ QEC rounds, it would flip the recorded state of the purple logical without triggering any detectors. However, with $d$ rounds of syndrome extraction, flipping the state of the purple logical with an undetectable error requires $d$ measurement errors. The same is the case for the corner movement step (\Cref{fig:LHTrans_d} and \Cref{fig:LHDW_d}), since the green logical string is updated by the new stabiliser measurements. \Cref{fig:ST_diagrams} provides a visualisation of this point. The movements of the logical corners are shown to be separated in time by $d$ rounds of error-correction, thereby maintaining minimum separation between boundaries of the same type.

For patch shrinking (\Cref{fig:LHTrans_e}, \Cref{fig:LHDW_e} and \ref{fig:LHDW_h}), we only require one round of syndrome extraction even though the logical strings are changed. This is because we can form detectors from the single-qubit measurements that shrink the patch. For example, in \Cref{fig:LHDW_e}, if four qubits were previously in the support of an $X$-type stabiliser, then the product of the single-qubit $X$ measurements on these qubits should (in the absence of noise) be the same as the previously-measured value of the stabiliser. Moreover, note that that stabiliser was measured for $d$ QEC rounds previously (during corner movement, \Cref{fig:LHDW_d}). Hence, a flip in the measurement outcome of the purple qubit in \Cref{fig:LHDW_e} would be detectable via this method, and a string of at least $d$ such errors would be required for the logical error to be undetectable.

Finally, we also need $d$ rounds of syndrome extraction at the patch extension step shown in \Cref{fig:LHDW_g}, even though the measurements are not used to update any logical operator. With $k$ QEC rounds, there are errors with $k+1$ fault locations that, under phenomenological noise, produce a logical error and no detector flips. For example, consider a $Z$ error occurring on qubit $(1,9)$ before the syndrome extraction rounds in \Cref{fig:LHDW_g}. This error flips the green logical string. Furthermore, since the first stabiliser measurement at $(2,10)$ is non-deterministic, no detector is flipped by this error at that location; it could only be picked up through measurements of the stabiliser at $(0,8)$. If $k$ measurement errors also occur at $(0,8)$, no detector will be flipped. Then, after the patch shrinking step shown in \Cref{fig:LHDW_h}, the $Z$ error is undetectable. Hence, we need $k\geq d-1$ to maintain the distance through this step and, in fact, under phenomenological noise, $k=d-1$ suffices. However, under circuit-level noise, at least $d$ rounds are required, as ``hook errors'' can effectively produce both a measurement error and a data qubit error from only a single Pauli error (see \Cref{sec:Stab_schedules} and \Cref{app:hook_error}).

The above method therefore involves $3d + 3$ rounds of syndrome extraction, not including those before and after the procedure is performed: $3d$ rounds during the three patch extension stages, $2$ further rounds during the patch shrinking steps and $1$ more after the first $SWAP$ step (\Cref{fig:LHDW_i}). We point out that the literature usually proposes $3d + 2$ rounds of syndrome extraction, since the syndrome extraction round between the two $SWAP$ layers is omitted; see e.g. \cite{Bombin2021}. While the overall effective distance is unchanged by omitting this step, the idling time of the qubits between stabiliser measurements is increased: we leave them idle for two $SWAP$ layers and $4$ $CX$/$CZ$ layers. If $SWAP$ is not a native gate, the number of layers between syndrome extraction increases further, which adds substantial noise. By inserting a round of syndrome extraction between the two $SWAP$ layers, we obtain a circuit with lower logical failure probability.

\subsection{Stabiliser measurement schedules for maintaining distance}\label{sec:Stab_schedules}

\begin{figure}[ht]
\centering
\includegraphics[width=0.75\textwidth]{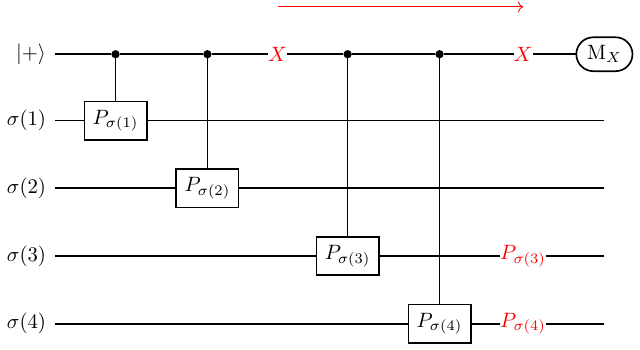}
    \caption{A hook error: one error on the auxiliary qubit spreads to two data qubits. If these two data qubit errors align with the logical operator, this reduces the effective code distance.}
    \label{fig:error_spreading}
\end{figure}

We now describe the scheduling used for syndrome extraction at each step in the logical Hadamard procedures. Consider an arbitrary stabiliser $g = P_1\cdots P_m$ where each $P_j$ is a weight-one Pauli defined on a unique qubit. Let us call the following circuit the auxiliary syndrome extraction circuit (\cite[Section 2]{tangled_schedules}) of the stabiliser $g$ under scheduling $P_{\sigma(1)},\cdots,P_{\sigma(m)}$ where $\sigma\in S_m$ is a permutation:
\begin{itemize}
    \item initialise an auxiliary qubit in the $|+\rangle$ state,
    \item sequentially apply the entangling gates controlled-$P_{\sigma(1)}$, controlled-$P_{\sigma(2)},$ ..., controlled-$P_{\sigma(m)}$, and optionally insert additional idling layers,
    \item measure the auxiliary qubit in the $X$ basis.
\end{itemize}
When one constructs a syndrome extraction circuit that measures a set of stabilisers simultaneously and independently, one usually constructs the above auxiliary circuits for each stabiliser, parallelises these circuits, verifies they form valid syndrome extraction circuits (see e.g. \cite[Figure 15]{LitOpp}), and finally compiles the obtained circuits to the desired native gate-set. The aim is for the effective distance of the code to remain unchanged after the introduction and compilation of these auxiliary circuits. For a more detailed explanation of auxiliary syndrome extraction circuits and scheduling rules, see \cite[Section 2]{tangled_schedules}.

\Cref{fig:error_spreading} shows the most damaging error possible under circuit-level noise when measuring a four-qubit stabiliser. When an $X$ error occurs on the auxiliary qubit between the second and third entangling gates of the auxiliary circuit, the error spreads to two data qubits. The remaining $X$ error on the auxiliary qubit is not flagged by the measurement, so the two data-qubit errors are only identified by other stabiliser measurements (that anticommute with this weight-two error) in either the same or the next QEC round, depending on the scheduling. We call this a ``hook error'', first observed in \cite{Dennis-Kitaev-Landahl}, where a single auxiliary qubit error becomes two data qubit errors. This could reduce the effective code distance if these two data qubit errors both form part of a minimal-weight logical string. In this case, only $\sim d/2$ faults are required to implement a logical error. This type of ``bad hook'' error should be avoided when constructing a syndrome extraction circuit for each step of the logical Hadamard procedure.

\begin{figure}
     \centering
     \begin{subfigure}[b]{0.5\textwidth}
         \centering
         \includegraphics[width=1\textwidth]{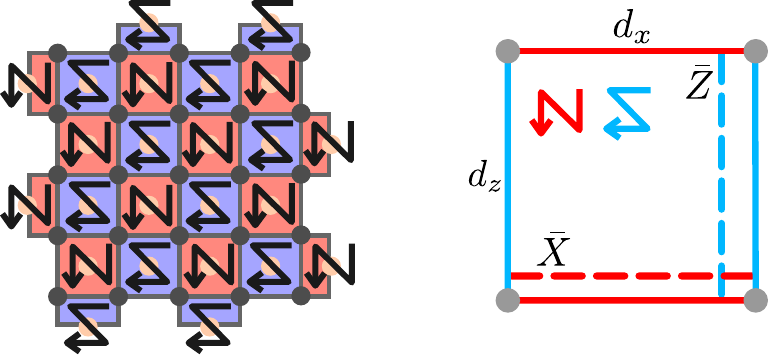}
         \caption{}
         \label{fig:scheduled_patch_0}
     \end{subfigure}\\
     \begin{subfigure}[b]{0.45\textwidth}
         \centering
         \includegraphics[width=1\textwidth]{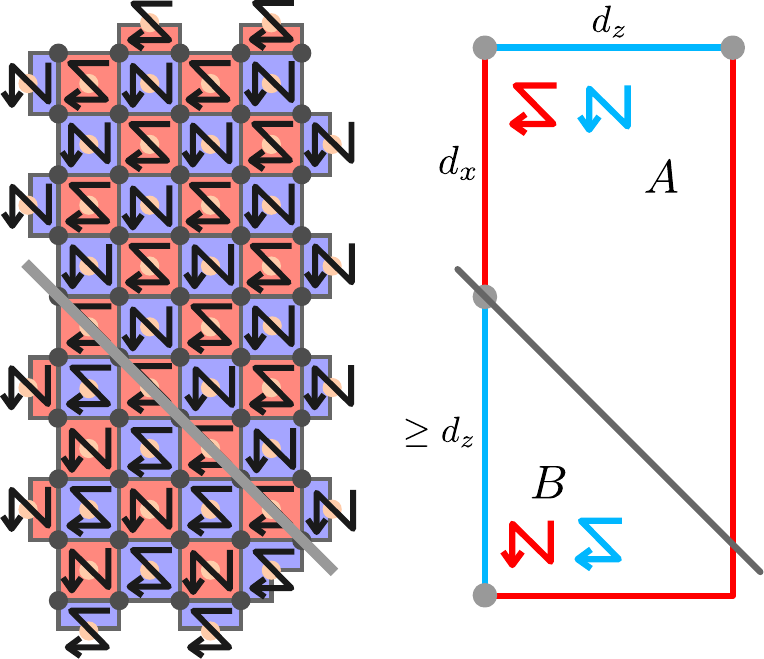}
         \caption{}
         \label{fig:scheduled_patch_1}
    \end{subfigure}\hfill
    \begin{subfigure}[b]{0.45\textwidth}
        \centering
        \includegraphics[width=\textwidth]{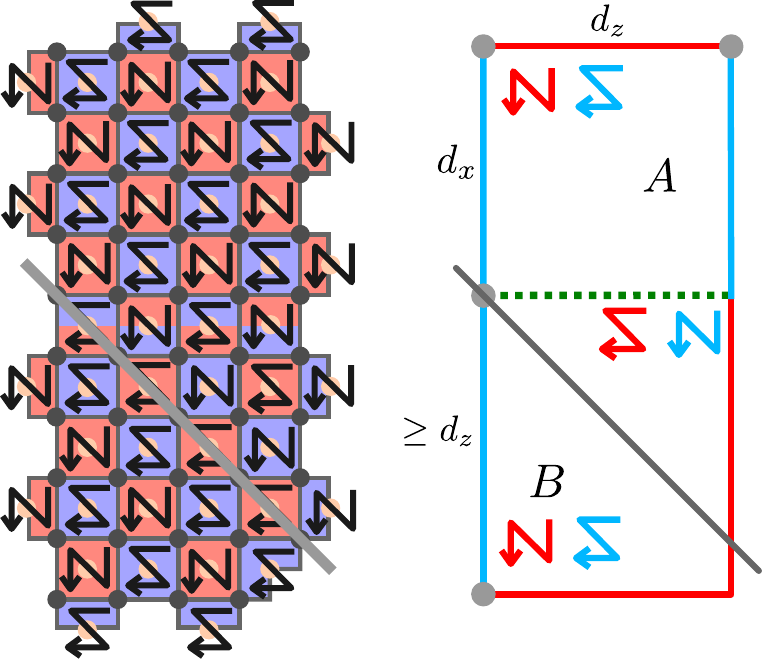}
        \caption{\label{fig:scheduled_patch_2}}
    \end{subfigure}
        \caption{(a) Left: Square, distance-$5$ surface code patch with plaquettes measured using the scheduling indicated. $Z$-type plaquettes are measured with a ``flipped-Z''-shaped schedule and $X$-type plaquettes with an ``N''-shaped schedule. Where the schedule arrow passes through locations at which there are no qubits (i.e. for boundary plaquettes), the circuit idles. For example, the entangling gates for the left-boundary, weight-$2$ $X$- and the bottom-boundary, weight-$2$ $Z$ stabilisers are performed in the $1$\textsuperscript{st} and $2$\textsuperscript{nd} layers. On the other hand, those for the top-boundary, weight-$2$ $Z$ and right-boundary, weight-$2$ $X$ stabilisers are performed in the $3$\textsuperscript{rd} and $4$\textsuperscript{th} layers. Right: Schematic illustration of the same patch with scheduling of $Z$ ($X$) plaquettes indicated with blue (red) arrows. Boundaries are coloured such that logical $Z$ ($X$) strings of lengths $d_z$ ($d_x$) can run along the blue (red) boundaries. Logical corners are indicated by grey dots and logical $\bar{Z}$ ($\bar{X}$) string representatives by blue (red) dashed lines. (b) Planar code with three logical corners on the left and one on the top-right. Left: the distance-5 patch with the scheduling of each plaquette labelled. The scheduling differs either side of the grey line. Right: a schematic illustration of the same patch. Sub-regions above (below) the grey line are labelled $A$ ($B$). The same scheduling maintains the effective distances in case the vertical blue boundary is chosen to be longer ($\geq d_z$) than that shown on the left. (c) The patch scheduling for the same patch in (b) but with a domain wall (dotted green line) inserted into the patch. The scheduling pattern in sub-regions $A$ and $B$ are the same but interchanged relative to the plaquette type ($X$ or $Z$) above the domain wall.}
        \label{fig:scheduled_patches}
\end{figure}

There exist schedulings of the rotated planar code of arbitrary $d_x\times d_z$ size that are devoid of bad hooks, such as that shown in \Cref{fig:scheduled_patch_0} for a square patch (see e.g.~\cite{Tomita_2014}). In the figure, the arrows in each plaquette trace out a path indicating the permutation $\sigma$; they begin at $\sigma(1)$ and terminate at $\sigma(4)$. The $X$ stabilisers follow an ``N''-shaped scheduling while the $Z$ stabilisers follow a ``flipped-Z''-shaped scheduling. These are sensible choices as they spread errors from the auxiliary qubits in a direction perpendicular to the corresponding logical strings. Namely, when measuring an $X$-type stabiliser, the hook error shown in \Cref{fig:error_spreading} results in $X$ errors on a pair of vertically adjacent data qubits. Since the minimal weight logical $X$ string is horizontal and has weight $d_x$ (see the logical string $\bar{X}$ in \Cref{fig:scheduled_patch_0}), this does not reduce the effective $X$-distance of the code patch, i.e. it is equal to the horizontal length of the patch: $d_x$. Note also that if this hook error occurs in the bulk of the patch, the outcome of the $Z$ stabiliser above the $X$ stabiliser is flipped in the same QEC round, while the $Z$ stabiliser below it will be flipped in the next QEC round. 

Similarly, an error from a $Z$-auxiliary qubit can spread to a pair of horizontally adjacent data qubits, while the direction of the minimal (i.e. $d_z$) weight logical $Z$ string is vertical. Therefore, the effective $Z$-distance is the vertical length of the patch, $d_z$. Furthermore, note that the $X$ stabiliser to the right is flipped in the same QEC round, while the one to the left is flipped in the next QEC round.

When implementing the logical Hadamard gate, we used two non-standard surface code patches (see \Cref{fig:LHTrans_c}--\ref{fig:LHTrans_d} and \Cref{fig:LHDW_c}--\ref{fig:LHDW_d}). Below we consider the case of the first non-standard patch in the transversal Hadamard method: \Cref{fig:LHTrans_c} (the others being similar), where we used a patch of size $d_z\times(d_x+d_z)$  -- this is $d\times(2d)$ in the square-shaped patch case. This patch has three logical corners located on its left-hand side boundary, and another at the top of its right-hand boundary. We detail the proposed scheduling in \Cref{fig:scheduled_patch_1} that maintains the effective $X$- and $Z$-distances as $d_x$ and $d_z$, respectively. We note that the blue vertical boundary can be chosen to be $\geq d_z$, in which case the proposed schedule still maintains the effective distances $d_x$ and $d_z$. After the application of the transversal Hadamard, the directions of $X$ and $Z$ logical strings are interchanged, so we interchange the scheduling on $X$ and $Z$ plaquettes: $X$ plaquettes follow a ``flipped-Z''-shaped scheduling while $Z$ plaquettes follow an ``N''-shaped scheduling. 

After performing patch extension (\Cref{fig:LHTrans_c}), we draw a grey line with gradient $-1$ from the logical corner half-way up the left boundary, as shown in \Cref{fig:scheduled_patch_1}. We define sub-region $A$ as all plaquettes whose auxiliary qubit lies above or on the grey line shown, and sub-region $B$ as all plaquettes whose auxiliary qubit lies below the grey line. If the stabiliser is in sub-region $A$, we do not change the scheduling. However, if it is in sub-region $B$, we once again interchange the schedules to a ``flipped-Z''-shaped scheduling for $Z$ plaquettes and an ``N''-shaped scheduling for $X$ plaquettes. It is straightforward to see that this scheduling defines a valid auxiliary syndrome extraction circuit for the patch's stabilisers, as it satisfies the conditions (a')-(b') from \cite[Section 2]{tangled_schedules} (alternatively, see \cite[Figure 15]{LitOpp}).

\begin{figure}
    \centering
    \begin{subfigure}[b]{0.3\textwidth}
        \centering
        \includegraphics[width=\textwidth]{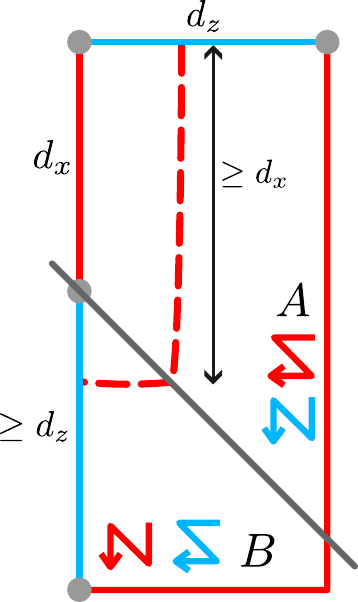}
        \caption{\label{fig:dx_distance_nonstandard_patch}}
    \end{subfigure}
    \hfill \begin{subfigure}[b]{0.3\textwidth}
        \centering
        \includegraphics[width=\textwidth]{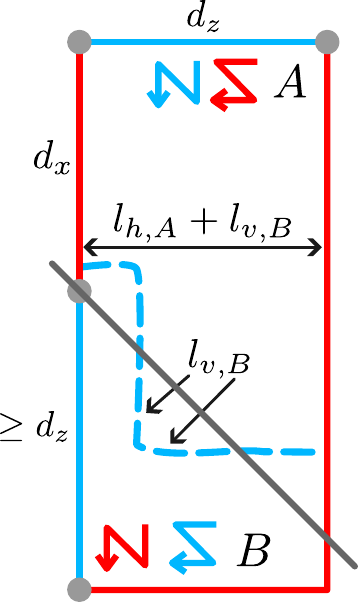}
        \caption{\label{fig:dz_distance_1}}
    \end{subfigure}
    \hfill \begin{subfigure}[b]{0.3\textwidth}
        \centering
        \includegraphics[width=\textwidth]{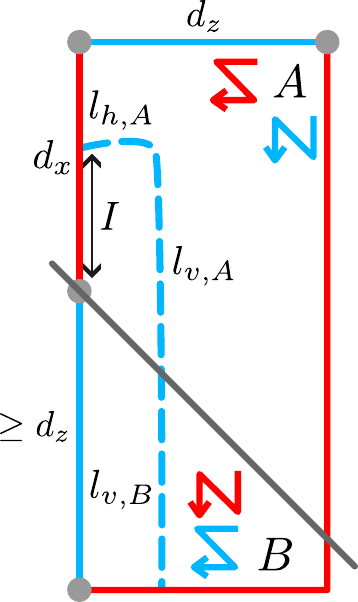}
        \caption{\label{fig:dz_distance_2}}
    \end{subfigure}
    \caption{Logical $X$ ($Z$) error strings (shown by dashed lines) for the non-standard rotated planar code patch shown in \Cref{fig:scheduled_patch_1} have effective weights at least as large as $d_x$ ($d_z$). In (a), the logical $X$ string is shown to require a vertical length through $A$ at least as large as $d_x$. Hence its effective weight is $\geq d_x$. In (b), we show that $l_{h,A} + l_{v,B} \geq d_z$ for a string that crosses the grey line an even number of times. In (c), we show a logical $Z$ string crossing the grey line an odd number of times. In this case, $l_{h,A} + l_{v,B} \geq l_{v,A} - I + l_{v,B} \geq d_z$ because of the length of the left-hand blue boundary. These logical $Z$ strings have effective weights $\geq d_z$.}\label{fig:logical_errors}
\end{figure}

We now explain why the choice of scheduling in \Cref{fig:scheduled_patch_1} maintains the effective $X$- and $Z$-distances. Let the ``effective weight'' of an error be the minimum number of fault locations required to create that error. Therefore, in the patch in \Cref{fig:scheduled_patch_0}, the effective weight of a pair of $Z$ errors on qubits $(x,y)$ and $(x+1,y)$ is $1$ for certain values of $x$ and $y$; it can arise from a single $X$ error on an auxiliary qubit. As for the effective $X$-distance of the patch shown in \Cref{fig:scheduled_patch_1}, let us consider first an $X$ logical string running between the two blue-coloured boundaries; see \Cref{fig:dx_distance_nonstandard_patch}. Since this red string must progress by at least $d_x$ steps vertically from the top boundary to half-way down the patch, the effective weight of any logical $X$ string must be at least $d_x$, owing to the scheduling in sub-region $A$. In fact, if the grey line were drawn horizontally, the $X$-distance would still remain $d_x$.

Now, we show that the choice of the grey line shown in \Cref{fig:scheduled_patch_1} maintains the effective $Z$-distance of the patch in \Cref{fig:scheduled_patch_1} as well. The desire is for the choice of sub-regions $A$ and $B$ to be such that the effective weight of any $Z$ logical string, i.e. $l_{h,A} + l_{v,B} + (l_{v,A} + l_{h,B})/2$, is at least $d_z$; here, $l_{h,A}$ is the horizontal distance traversed in sub-region $A$, $l_{v,B}$ is the vertical distance traversed in $B$, and so on. In fact we will show that $l_{h,A} + l_{v,B} \geq d_z$. With a $(-1)$-gradient boundary line between $A$ and $B$, any string that remains in $A$ clearly has $l_{h,A} = d_z$. We claim that any string that crosses $n>0$ times between $A$ and $B$ has $l_{h,A} + l_{v,B} \geq d_z$. To see this, we consider two sub-cases corresponding to $n$ even and $n$ odd, depicted in \Cref{fig:dz_distance_1} and \Cref{fig:dz_distance_2} respectively. Note that, between adjacent intersection points, the vertical and horizontal distances traversed are equal, owing to the gradient of the boundary line. Therefore, on the one hand, whenever there are an even number of intersection points (so that the line terminates in $A$), we have $l_{v,B} = l_{h,B}$ (see \Cref{fig:dz_distance_1}). In this case, $l_{h,A} + l_{v,B} = l_{h,A} + l_{h,B} \geq d_z$, since the total horizontal distance traversed must be at least $d_z$. On the other hand, when there are an odd number of intersection points, $l_{h,A} + l_{v,B}$ is at least the length of the vertical blue boundary, which itself has length $\geq d_z$. This is because, $l_{h,A} = l_{v,A} - I$ where $I$ is the distance from the string's starting point to the logical corner (see \Cref{fig:dz_distance_2}). Meanwhile $l_{v,A} - I + l_{v,B}$ is the length of the left-hand blue boundary, which is at least $d_z$. Therefore, the effective weight of any logical $Z$ string is indeed at least $d_z$.


\section{$SWAP$-QEC round with four two-qubit gate layers}\label{sec:compilation}

In the previous sections, each QEC round during the execution of a logical Hadamard gate experiment can be performed by resetting some qubits, applying four layers of two-qubit gates ($CX$ or $CZ$), and finally measuring some qubits. However, when we shift the patch in the final two stages, we need an additional layer of $SWAP$ gates. Here, we demonstrate that, using circuit simplifications, we can execute the $SWAP$ layer together with the QEC round that follows it using just four two-qubit gate layers. 

We remark that a recent paper \cite{McEwenBaconGidney} demonstrated a similar circuit simplification result, which they called ``stepping'' circuits. They demonstrated via numerical simulations that the performance of these circuits is essentially the same as that of the standard syndrome extraction circuits. In this section, our contribution is to show how one can construct these ``stepping'' circuits from the standard ones using some circuit simplification ideas that can also be applied to other codes.

\begin{figure}[ht]
    \centering
    \begin{subfigure}[b]{0.5\textwidth}
        \centering
        \includegraphics[width=0.6\textwidth]{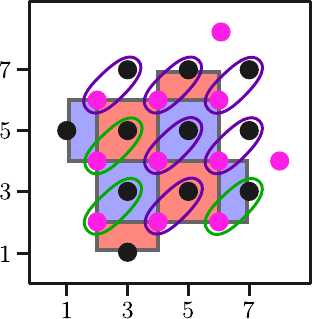}
        \caption{}\label{fig:wiggle_patch_1}
    \end{subfigure}\hfill
    \begin{subfigure}[b]{0.5\textwidth}
        \centering
        \includegraphics[width=0.6\textwidth]{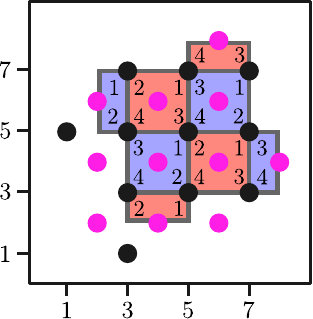}
        \caption{}\label{fig:wiggle_patch_2}
    \end{subfigure}
    \caption{A QEC round where we shift a $3\times 3$ patch in the north-east direction. The patch whose stabilisers we measured in the first QEC round is shown in (a), and the second QEC round's stabilisers are shown in (b). Pink/black dots are qubits that serve as data/auxiliary qubits for the QEC round in (a) and serve as auxiliary/data qubits for the round in (b). We swap the encircled pairs of qubits, with different procedures occurring for purple/green encircled qubits; see text for explanation. In (b), the numbers on the plaquettes indicate the scheduling in the auxiliary syndrome extraction circuits of the stabilisers for that QEC round.}\label{fig:Shifted}
\end{figure}

Now, we explain these circuit simplifications for the $SWAP$-QEC rounds. We pair the data qubits of the current QEC round with data qubits of the next QEC round as shown in \Cref{fig:Shifted} by green and purple encircled pairs. In the figure, we colour qubits pink or black. In \Cref{fig:wiggle_patch_1}, we show the stabilisers for the current QEC round (call it the (a)-round), where the pink qubits are the data qubits. In the next QEC round (the (b)-round), the stabilisers are shown in \Cref{fig:wiggle_patch_2} and the black qubits are the data qubits. To determine the circuit we enact, we first encircle all diagonally adjacent qubits as shown in \Cref{fig:wiggle_patch_1} (i.e. pair pink qubit at $(x,y)$ with black qubit at $(x+1,y+1)$). If the pink qubit in each circle is to be an auxiliary qubit in the (b)-round, we colour the circle purple; otherwise we colour it green. Note that those black qubits in purple circles are scheduled first in some auxiliary syndrome extraction circuit in the (b)-round (specifically the circuit including the pink auxiliary qubit with which it is swapped; see the numbering in \Cref{fig:wiggle_patch_2}), while the black qubits in the green circles are not. This alignment of the $SWAP$ gates with the first layer of entangling gates in the standard syndrome extraction circuit is crucial for the simplification of the $SWAP$-QEC round circuit.

\begin{figure}[ht]
\centering
\includegraphics[width=0.8\textwidth]{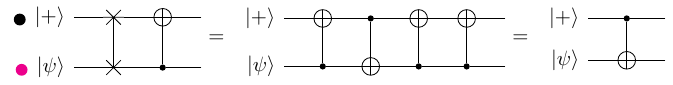}
\caption{Circuit simplification for the case when the stabiliser's first scheduled Pauli term is $X$, and we swap the data and auxiliary qubits. The $SWAP$ and the $CX$ gates can be simplified into just one entangling layer.}\label{fig:SWAP_and_first_CX_layer}
\end{figure}

\begin{figure}[ht]
    \centering
    \includegraphics[width=0.75\textwidth]{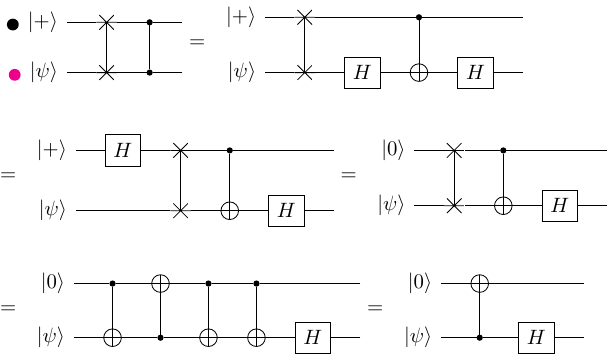}
    \caption{Circuit simplification for the case when the stabiliser's first scheduled Pauli term is $Z$, and we swap the data and auxiliary qubits. The qubit initialisation, the $SWAP$ and the $CZ$ gates can be simplified into just a $CX$ gate and a Hadamard gate. Note that the basis of initialisation is changed to $Z$.}\label{fig:SWAP_and_first_CZ_layer}
\end{figure}

\begin{figure}[ht]
    \centering
    \includegraphics[width=0.3\textwidth]{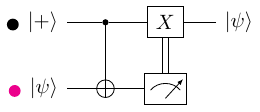}
    \caption{Circuit for teleportation.}\label{fig:Teleport}
\end{figure}

Now, consider first those qubits in purple circles. Since the first entangling gate in the syndrome extraction circuit for the (b)-round is enacted on the same two qubits that have just been swapped, we can simplify the qubit initialisation, the $SWAP$ and the first entangling gate of the auxiliary circuit as shown in \Cref{fig:SWAP_and_first_CX_layer} (when the auxiliary circuit is that of an $X$-type stabiliser) and \Cref{fig:SWAP_and_first_CZ_layer} ($Z$-type stabiliser). Note that the additional Hadamard gate in \Cref{fig:SWAP_and_first_CZ_layer} can be used to re-compile the rest of the $Z$-plaquette auxiliary syndrome extraction circuit to use $CX$ gates and $Z$ measurements instead of $CZ$ gates and $X$ measurements. The case of the green-encircled qubits is handled by qubit teleportation, as shown in \Cref{fig:Teleport}. After these, we continue with the auxiliary syndrome extraction circuits for the (b)-round. Note that the Pauli $X$ correction from \Cref{fig:Teleport} can be applied in software. If we do this, however, we are effectively tracking a data qubit $X$ error, hence, several $Z$ stabilisers' outcomes and possibly the logical observable are flipped. Therefore, we need to update the detectors and the observable to accommodate the teleportation measurement outcomes.

\begin{figure}[h!]
    \centering
    \includegraphics[width=0.5\textwidth]{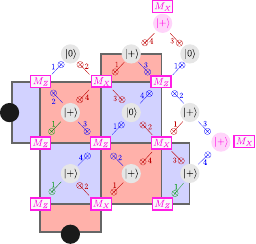}
    \caption{Full $SWAP$-QEC round compiled to initialisation and measurement in $X$ and $Z$ bases, and $CX$ gates. The layout is the same as shown in \Cref{fig:Shifted}. The plaquettes of the (a)-round are shown. Pink and black qubits are now indicated by circles showing initialisation basis and/or boxes showing the measurement basis used at the end of the circuit. Red $CX$ gates are used for measuring $X$-plaquettes in the (b)-round, blue ones for $Z$-plaquettes, and green ones are for teleportation. Layers in which each $CX$ is applied are indicated by numbers. All initialisations happen in layer $0$ and measurements in layer $5$.}\label{fig:SWAP_QEC_round}
\end{figure}

\Cref{fig:SWAP_QEC_round} illustrates how the $SWAP$-QEC round can be executed after the circuit simplifications are applied. In summary, we use the same number of layers as for a standard QEC round; however, because of teleportation, we need some extra $CX$ gates, initialisations and measurements. This, of course, adds additional noise to the qubits and thus, when evaluating it with Pauli noise (ignoring e.g. leakage), we would expect a somewhat larger logical failure probability. As the patch size grows, however, this becomes more negligible. The scheduling used in \Cref{fig:Shifted} is suitable for the first $SWAP$-QEC round of the logical Hadamard experiment. However, for the second $SWAP$-QEC round we have to reflect each plaquette's scheduling horizontally through a vertical line, since the $SWAP$'s direction now is north-west.

We close this section by pointing out that our circuit simplifications can be applied to codes other than the planar code. All we need to ensure is that the first layer of entangling gates aligns with the $SWAP$s we intend to perform on the code, and then the simplifications discussed above apply. For instance, consider the triangular colour code on a hexagonal lattice with one auxiliary qubit per hexagonal plaquette, placed at the centre of each hexagon~\cite{thomsen2022lowoverhead}. Recall that each hexagon corresponds to an $X$- and a $Z$-type stabiliser. Assume further that, for each $X$ stabiliser, the data qubit that is north of the auxiliary qubit is scheduled first, while for each $Z$ stabiliser, the data qubit that is south of the auxiliary qubit is scheduled first. Then, using the circuit identities described above, it is straightforward to construct a syndrome extraction circuit that swaps the roles of data and auxiliary qubits twice. Namely, we can combine one $SWAP$ layer with the circuit that measures the $X$ stabilisers, which therefore shifts the code in the north direction; and after that we can do the same while measuring the $Z$ stabilisers and shift the code back into its original position. This requires twelve two-qubit gate layers, two auxiliary qubit intitialisation layers, and two auxiliary qubit measurement layers -- the same as for a standard QEC round for this code.


\section{Simulation results}\label{sec:simulation}

Our numerics use the Python library \texttt{stim} \cite{stim} to prepare and sample the circuits, and \texttt{pymatching} \cite{pymatching} to decode the syndromes. All the circuits described below and used for our simulations are available in \cite{our_stim_circuits}. We compare the logical failure probability of the logical Hadamard experiment to that of the standard quantum memory experiment, which we performed on the same patch for the same number of QEC rounds. We now provide details of these simulations.

Recall that, in the standard quantum memory experiment for a CSS code, we do simulations for two different circuits: one for the $X$-memory case, and another for the $Z$-memory case. For $X$-memory, we initialise all data qubits in the $|+\rangle$ state, perform several QEC rounds of syndrome extraction (usually proportional to $d$), the first of which initialises the logical $|+_L\rangle$ state on the patch, and finally measure all data qubits in the $X$ basis. Based on the measurement outcomes, we form detectors \cite{stim}, we decode the syndrome using a decoding graph and finally, from the decoded data, we form our logical $X$ outcome. This circuit tests only for $Z$-type errors. The $Z$-memory case is very similar, the only difference being that we initialise/measure the data qubits in the $Z$ basis instead of $X$. The $Z$-memory circuit tests for $X$-type errors only. To obtain the logical failure probability $p_L^{QM}$ of the patch for the given number of QEC rounds, we sample both experiments, take the obtained probabilities $p_L^X$ and $p_L^Z$ from both, and combine them as 
\begin{equation}
    p_L^{QM} = p_L^X + p_L^Z - p_L^X \cdot p_L^Z;
\end{equation}
see \cite[Eq. 1]{GidneyHoneycomb}. This is the probability that \textit{either} a logical $X$ error or a logical $Z$ error occurs on the patch, i.e. that any logical failure occurs.

For the logical Hadamard case, we consider two experiments, which are natural modifications of the two quantum memory experiments. We call these circuits the \emph{$X$-to-$Z$} and \emph{$Z$-to-$X$ Hadamard experiments}, since the Hadamard gate interchanges the $X$ and $Z$ bases. For the $X$-to-$Z$ Hadamard experiment, we initialise the logical $|+_L\rangle$ state on the patch by initialising all data qubits in the $|+\rangle$ state and performing one QEC round on the initial patch, then we continue by performing the logical Hadamard gate (using one of the methods from \Cref{sec:explanatation}), and finally, after the last $SWAP$ layer, we perform one further QEC round then measure out the logical $Z$ operator by measuring all data qubits in the $Z$ basis. The $Z$-to-$X$ memory experiment is very similar, the only difference being that we initialise all data qubits in the $|0\rangle$ state in the beginning and measure them in the $X$ basis at the end. Note that both circuits include $2$ additional QEC rounds, i.e. in total $3d+5$, as compared to the counted $3d+3$ in \Cref{sec:distance_repeat_mmts}. This is because of the logical state initialisation and the logical operator readout, that do not count towards the logical Hadamard gate procedure during FTQC. To obtain the logical failure probability $p_L^{LH}$ of the patch under the logical Hadamard gate, we perform both experiments, take the obtained probabilities $p_L^{X\to Z}$ and $p_L^{Z\to X}$ and, as for the quantum memory case, we combine them as
\begin{equation}
    p_L^{LH} = p_L^{X\to Z} + p_L^{Z\to X} - p_L^{X\to Z} \cdot p_L^{Z\to X}.
\end{equation}

\begin{figure}[h!]
    \centering
    \begin{subfigure}[b]{0.45\textwidth}
        \centering
        \includegraphics[width=\textwidth]{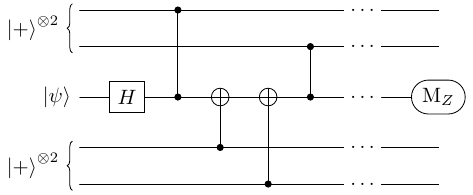}
        \caption{}
    \end{subfigure}\hfill
    \begin{subfigure}[b]{0.39\textwidth}
        \centering
        \includegraphics[width=\textwidth]{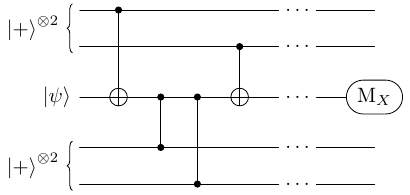}
        \caption{}
    \end{subfigure}\\
    \begin{subfigure}[b]{0.5\textwidth}
        \includegraphics[width=\textwidth]{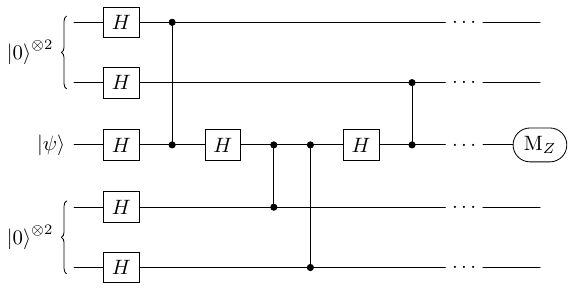}
        \caption{\label{fig:circuit_equiv_c}}
    \end{subfigure}
    \caption{Entangling gates and measurements enacted on one of the data qubits (e.g. qubit $(3,17)$) in the surface code patches during the first five steps of the transversal Hadamard (a) and domain wall (b) procedures (i.e. up to the patch-shrinking step, see \Cref{fig:log_had_transversal_patches} and \ref{fig:log_had_dom_wall_patches}). Compiling both to the native gate set of $RZ$, $MZ$, $H$ and $CZ$ we arrive at the circuit in (c). For qubits in the bottom half of the extended patches, the circuits applied to them are already equivalent in the two procedures.}\label{fig:circuit_equiv}
\end{figure}

Each of the above-described circuits were compiled using the following gates: $Z$-basis reset ($RZ$), $Z$-basis measurement ($MZ$), Hadamard gate ($H$), and controlled-$Z$ gate ($CZ$). We point out that, under this choice of native gates, the compiled circuits for the two logical Hadamard procedures (in either $X$-to-$Z$ or $Z$-to-$X$ experiments) are equivalent. Indeed, as can be seen from \Cref{fig:ST_diagrams}, the two procedures, which differ only above the line $y=10$ (see \Cref{fig:log_had_transversal_patches} and \ref{fig:log_had_dom_wall_patches}), are equivalent upon conjugating the circuit acting on the patch above $y=10$ with Hadamard gates. Considering the transversal Hadamard procedure, this effectively pushes the time-like domain wall through the circuit, leaving the space-like domain wall separating the two halves of the rectangular patch, and therefore mapping the circuit to that of the domain wall patch deformation procedure. Compiling the circuit to a given choice of native gates can effectively perform this conjugation, leaving two compiled circuits that are precisely the same. In \Cref{fig:circuit_equiv}, we present the circuits for steps (a)-(e) of \Cref{fig:log_had_transversal_patches} and \Cref{fig:log_had_dom_wall_patches} for a single data qubit in state $\ket{\psi}$ in the bulk of the lattice and hence in the support of four stabilisers. This qubit undergoes four entangling gates of types $CX$ or $CZ$ per QEC round and, in the patch-shrinking step, is measured in the $Z$ or $X$ basis for the transversal Hadamard or domain wall procedure, respectively. In \Cref{fig:circuit_equiv_c} we present the same circuits compiled to the native gate set described above ($RX$ is replaced with $RZ\cdot H$, $CX$ is replaced with $H\cdot CZ\cdot H$, $H^2$ is removed from the circuit, $MX = H\cdot MZ$, etc.). We note that compiling the circuits to $RX$, $CX$, $H$ and $MX$ may also produce equivalent circuits for the two procedures depending on the precise compilation rules used.

To simulate the noisy circuits, we used a popular noise model which captures the idea that, on superconducting hardware, the noisiest operations are typically two-qubit gates and measurements. This noise model was described e.g. in \cite{tangled_schedules} (see also \cite{Paler2022PipelinedCM, barber2023realtime}); it is parametrised by $p$, the physical error rate, and applies the following noise operations:
\begin{itemize}
    \item a two-qubit depolarising channel with strength $p$ after each entangling gate,
    \item a classical flip of each measurement outcome with probability $p$,
    \item a one-qubit depolarising channel with strength $p/10$ after each Hadamard, reset or measurement gate,
    \item a one-qubit depolarising channel with strength $p/10$ on each idling qubit in each layer.
\end{itemize}

\begin{figure}[h!]
    \centering
    \begin{subfigure}[b]{0.46\textwidth}
        \includegraphics[width=\textwidth]{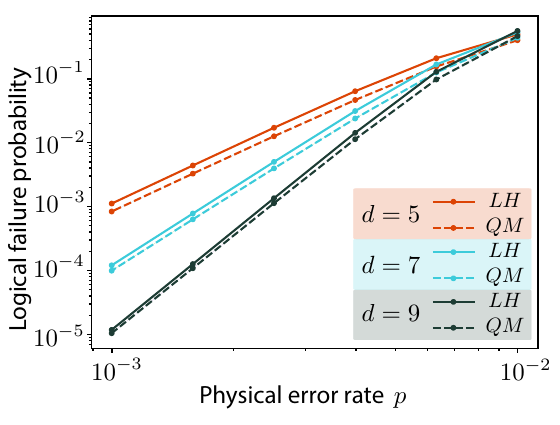}
        \caption{\label{fig:log_fail_plot}}
        \end{subfigure}\hfill
    \begin{subfigure}[b]{0.54\textwidth}
        \includegraphics[width=\textwidth]{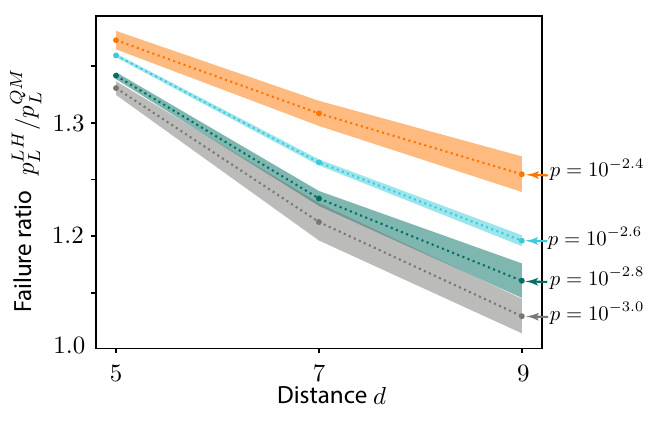}
        \caption{\label{fig:ratio_plot}}
    \end{subfigure}
    \caption{Simulation data for the logical failure probabilities $p_L^{LH}$ and $p_L^{QM}$ of the logical Hadamard gate and quantum memory experiments, respectively. (a) Logical failure probabilities are plotted for a range of physical error rates. (b) Ratio of logical failure probabilities, $p_L^{LH} / p_L^{QM}$, for four different physical noise rates, with confidence intervals of $95\%$ shown}
\end{figure}

Our simulation data are summarised in \Cref{fig:log_fail_plot} where we show simulation results for distances $5$, $7$ and $9$. We sampled $n=10^7$ shots for physical error rates $p\in\{10^{-2.4},10^{-2.2},10^{-2}\}$, and $n=10^9$ shots for $p\in\{10^{-3},10^{-2.8},10^{-2.6}\}$, except for the case when $p=10^{-3}, d=9$ where we sampled $n=10^{10}$ shots. As a result, the standard error of the mean error bars are too small to show in \Cref{fig:log_fail_plot}. It is clearly visible that the logical Hadamard gate's QEC performance is slightly worse than that of quantum memory. However, as the distance increases or the physical error rate decreases, the two performances become closer. In order to quantify this, we plot the ratio of the sampled logical failure probabilities, ${p_L^{LH}}/{p_L^{QM}}$, in \Cref{fig:ratio_plot} for the four smallest physical error rates. Shaded regions around each line correspond to $95\%$ confidence intervals. These are given by the ratios of extremal values of the $\sqrt{95}\% = 97.47\%$ confidence intervals for the two sampled logical error probabilities (each confidence interval being $p_L \pm 2.24\cdot\sigma/\sqrt{n}$, for $\sigma$ the sample standard deviation and $n$ the number of shots). Note that the width of the confidence interval depends on the number of shots, the logical failure probabilities, and how close the two sampled probabilities are. Therefore, we conclude that the logical Hadamard gate's QEC performance is very similar to that of quantum memory for an equivalent number of rounds.

\section{Conclusion}\label{sec:conclusion}

In this paper, we presented explicit circuit constructions for a logical unitary gate on a rotated planar code, and evaluated its performance under Pauli circuit-level noise. We considered two ways to perform the logical Hadamard gate, and explained in detail why they work. We also gave insight into the number of necessary QEC rounds for each intermediate step. For the non-standard planar code patches that the logical Hadamard procedure requires, we constructed syndrome extraction circuits that maintain the effective distance of the patch. We also explained how we can shift the patch in a given direction using only four two-qubit gate layers, an explanation that is applicable to other codes as well. As a byproduct, we gave an alternative gate-based derivation for the ``stepping'' circuits presented in \cite{McEwenBaconGidney}.

With numerical simulations, we found only small differences between the QEC performance of the logical Hadamard gate and that of a quantum memory experiment run for the same number of QEC rounds. These are likely due to the rectangular patches involved in the logical Hadamard procedures (see \Cref{fig:log_had_transversal_patches} and \Cref{fig:log_had_dom_wall_patches}) introducing more pathways for logical failures. However, note that the same rectangular patches allow the codes to tolerate other higher-weight errors. Since the differences between logical Hadamard and quantum memory performance become small for low physical error rates and large distances, they will likely be negligible for large-scale, fault-tolerant quantum computation, where low logical failure probabilities will be required. However, one must take into account the potential extra time and spatial overheads that are incurred by implementing unitary logical gates directly, when compared with commuting these gates through the remaining circuit (Pauli-based computation)~\cite{GoSc,blunt2023compilation}. We also point out that, during the logical Hadamard gate procedure, unlike in quantum memory, each data qubit of the original patch is reset once (see \Cref{fig:LHDW_f}), and hence we expect the logical Hadamard gate to be more resilient to leakage. Of course, the $SWAP$-QEC procedures of \Cref{sec:compilation} could be used in the regular quantum memory experiment to reduce leakage. Further work is needed to determine how the fidelity of the logical Hadamard would compare to quantum memory in noise models that include leakage.

Several other proposed logical operations remain under-explored in the circuit-level paradigm. The next natural candidate would be the logical $S$ gate on the rotated planar code, for which there are two existing proposals. One was presented in \cite{Bombin2021}, which performs the gate via patch-deformation; the other procedure prepares an auxiliary patch in the logical $Y$ eigenstate (see e.g. \cite{Gidney2023InplaceAT} for this), performs a joint logical $ZZ$-measurement and finally measures the auxiliary patch in $X$ basis (see e.g. \cite{GoSc}). It would be interesting to compare these two methods and assess which one performs better under circuit-level noise. Another natural next step would be to assess the QEC performance of the logical $CX$ gate (which can be performed via lattice surgery), or that of magic state distillation~\cite{Litinski_2019}, under circuit-level noise. Of course, it would also be fruitful to go beyond the planar code in simulations. Performing similar investigations on other QEC codes would allow for comparisons between very different fault-tolerant strategies under realistic noise models.

\section*{Note added}

After submitting this manuscript one reviewer noted that the supplementary material for Ref.~\cite{gidney2023yokedsurfacecodes} included data from simulated surface code rotation experiments, though the relevant results were not presented in the paper itself.


\bibliography{main}

\appendix

\newpage

\section{Logical Hadamard as logical corner braiding}\label{app:Corner_Braiding}

\begin{figure}[h!]
    \centering
    \begin{subfigure}[b]{0.3\textwidth}
        \centering
        \includegraphics[width=0.75\textwidth]{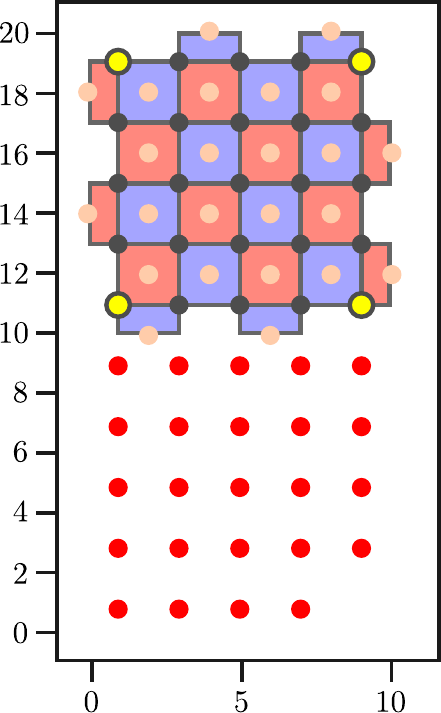}
        \caption{\label{fig:app_corner_move_a}}
    \end{subfigure}\hfill
    \begin{subfigure}[b]{0.3\textwidth}
        \centering
        \includegraphics[width=0.75\textwidth]{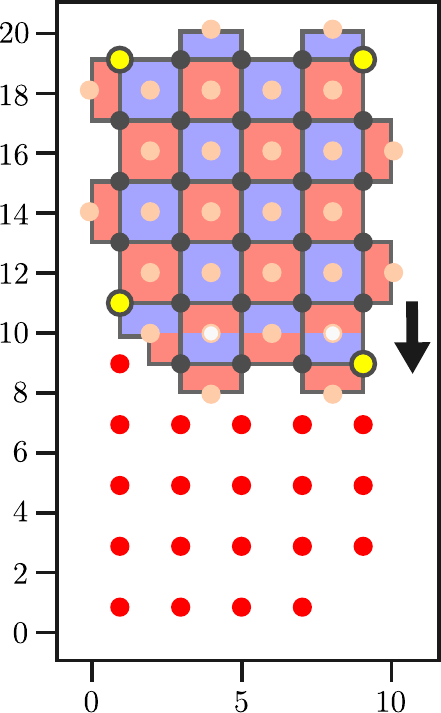}
        \caption{\label{fig:app_corner_move_b}}
    \end{subfigure}\hfill
    \begin{subfigure}[b]{0.3\textwidth}
        \centering
        \includegraphics[width=0.75\textwidth]{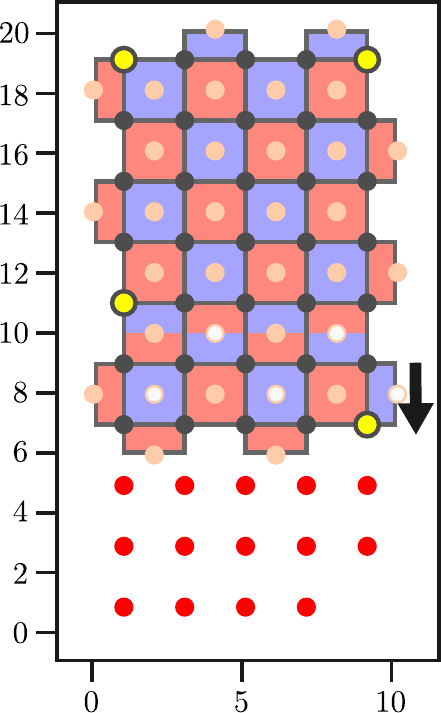}
        \caption{\label{fig:app_corner_move_c}}
    \end{subfigure}\\
    \begin{subfigure}[b]{0.3\textwidth}
        \centering
        \includegraphics[width=0.75\textwidth]{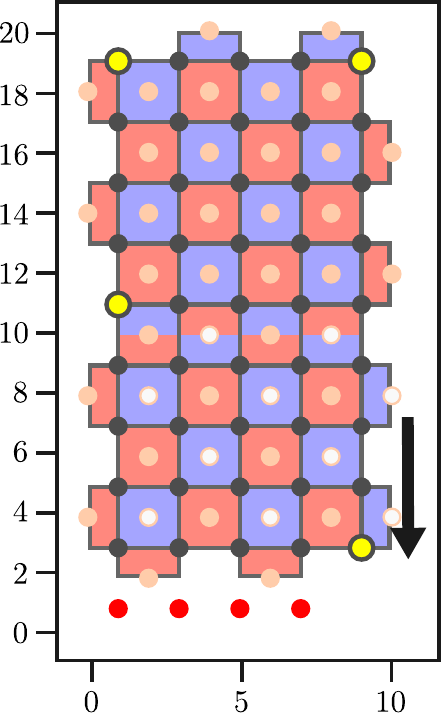}
        \caption{\label{fig:app_corner_move_d}}
    \end{subfigure}\hfill
    \begin{subfigure}[b]{0.3\textwidth}
        \centering
        \includegraphics[width=0.75\textwidth]{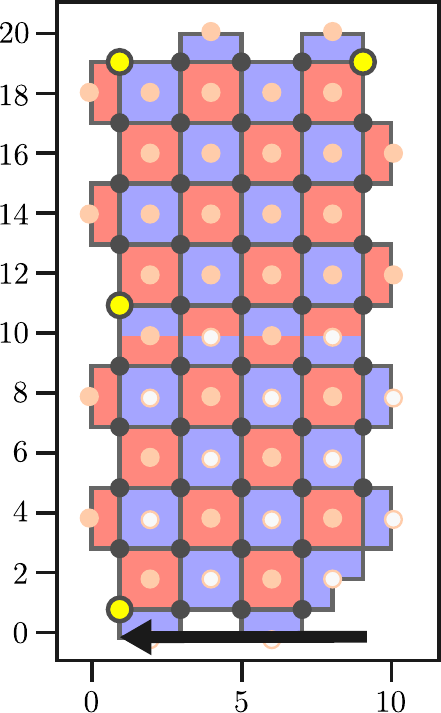}
        \caption{\label{fig:app_corner_move_e}}
    \end{subfigure}\hfill
    \begin{subfigure}[b]{0.3\textwidth}
        \centering
        \includegraphics[width=0.75\textwidth]{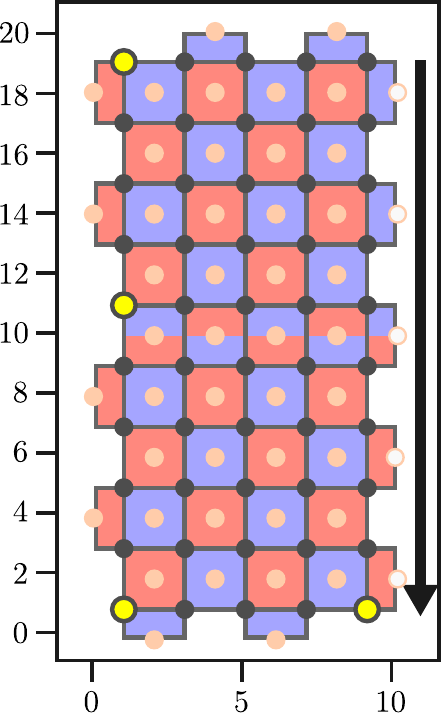}
        \caption{\label{fig:app_corner_move_f}}
    \end{subfigure}
    \caption{The patch extension measurements of \Cref{fig:LHDW_c} braid the logical corners (shown as yellow vertices), moving the bottom-right logical corner to a new location in the bottom-left of the extended patch. We show this by considering each row of stabiliser measurements. (a) - (d): The stabiliser measurements move the logical corner down. (e): These measurements move the logical corner to the bottom-left of the patch. (f): The measurements of \Cref{fig:LHDW_d} move the right-hand corner down the patch.}
    \label{fig:corner_move_1}
\end{figure}

\begin{figure}[h!]
    \centering
    \begin{subfigure}[b]{0.22\textwidth}
        \centering
        \includegraphics[width=\textwidth]{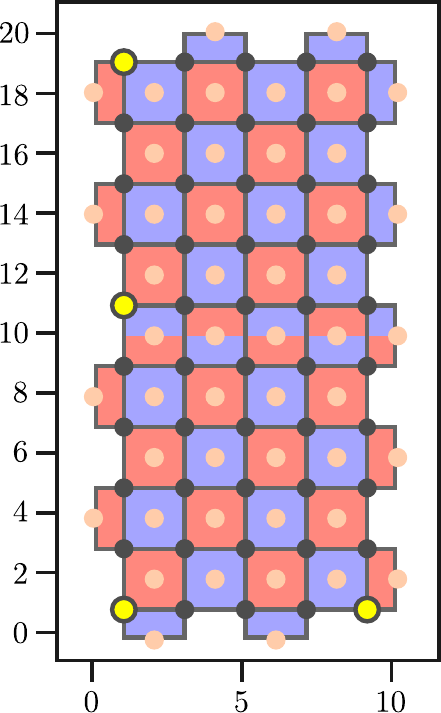}
        \caption{\label{fig:app_corner_move_2_a}}
    \end{subfigure}\hfill
    \begin{subfigure}[b]{0.22\textwidth}
        \centering
        \includegraphics[width=\textwidth]{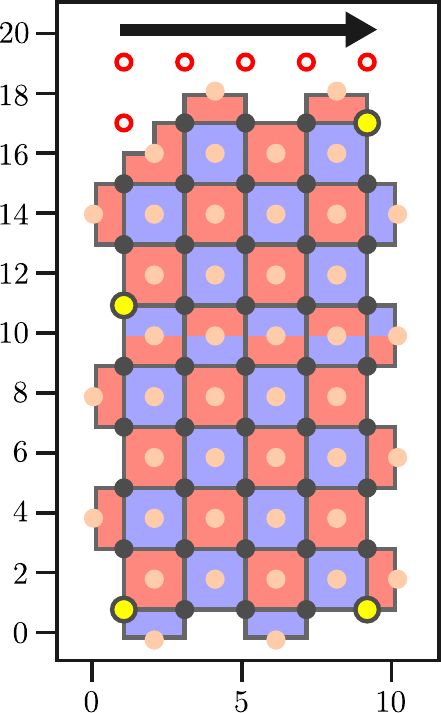}
        \caption{\label{fig:app_corner_move_2_b}}
    \end{subfigure}\hfill
    \begin{subfigure}[b]{0.22\textwidth}
        \centering
        \includegraphics[width=\textwidth]{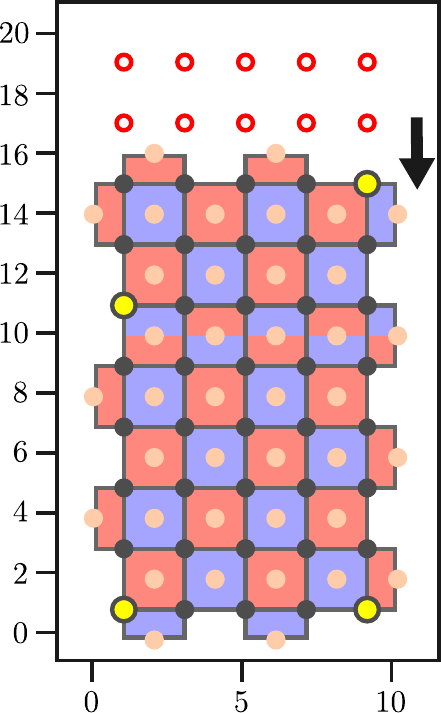}
        \caption{\label{fig:app_corner_move_2_c}}
    \end{subfigure}\hfill
    \begin{subfigure}[b]{0.22\textwidth}
        \centering
        \includegraphics[width=\textwidth]{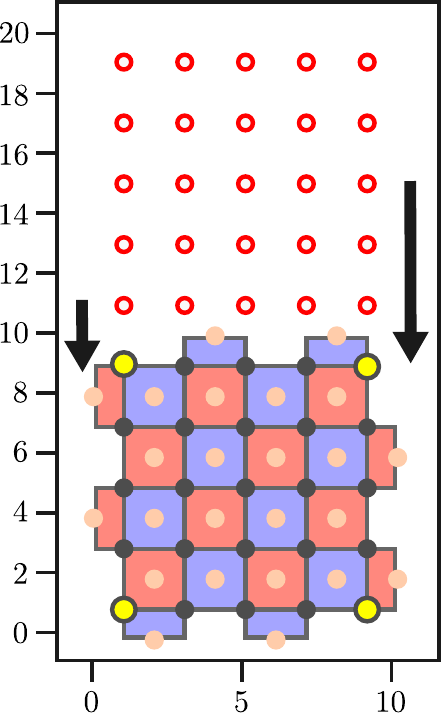}
        \caption{\label{fig:app_corner_move_2_d}}
    \end{subfigure}\hfill
    \caption{Patch-shrinking measurements of \Cref{fig:LHDW_e} can be interpreted as moving the logical corner from the top-left to the right ((a)--(b)) and downwards ((c)--(d)), and the logical corner at $(1,11)$ down by one data qubit ((d)).}
    \label{fig:corner_move_2}
\end{figure}

Here, we provide a topological interpretation of the logical Hadamard procedures of \Cref{sec:explanatation} via braiding logical corners, which is known to implement a Hadamard gate on the logical qubit~\cite{Brown_2017,Bombin_2010,landahl2023logical}. Our procedures effectively enact this braid, as we will show. Note that there is a correspondence between logical corners of a surface code patch and twist defects -- objects that appear at the ends of domain walls~\cite{Kesselring_2018}. Hence, we can equivalently think about braiding twists. We present the details for the domain wall patch deformation procedure. In \Cref{fig:corner_move_1}, we display two patch deformation steps from \Cref{fig:log_had_dom_wall_patches} and show, by considering subsequent rows of stabiliser measurements, how they can be interpreted as corner movements. In \Cref{fig:corner_move_2}, we similarly show how one can interpret the patch shrinking measurements of \Cref{fig:LHDW_e} as the movement of two logical corners. After this, the remaining steps in the logical Hadamard procedure simply move the patch back to its original location. They do not perform any exchange of the logical corners and so no logical operation is performed by them.

\begin{figure}[h!]
    \centering
    \includegraphics[width=\textwidth]{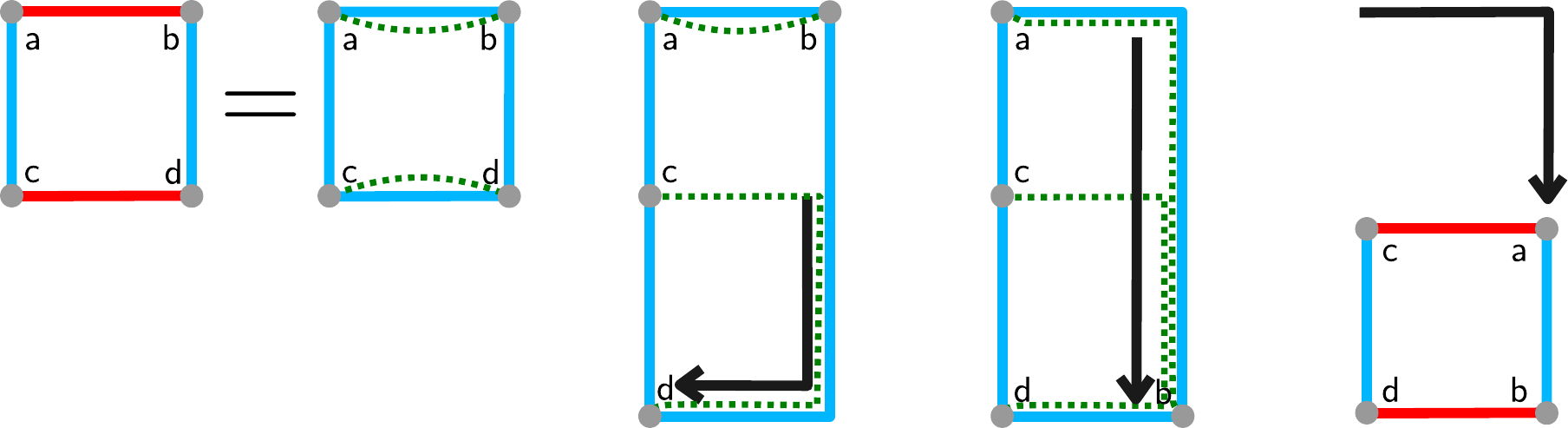}
    \caption{Schematic illustration of the moving of the logical corners shown in \Cref{fig:corner_move_1} and \ref{fig:corner_move_2}. We first illustrate a red-coloured boundary equivalently as a blue-coloured boundary and a domain wall superimposed. At the end of the procedure, the boundary colours are restored. Two domain walls superimposed are equivalent to no domain wall. The result of the procedure is the rotation of the logical corners.}
    \label{fig:twist_braid_schem}
\end{figure}

\begin{figure}[h!]
\centering
    \begin{subfigure}[b]{0.25\textwidth}
        \centering
        \includegraphics[width=\textwidth]{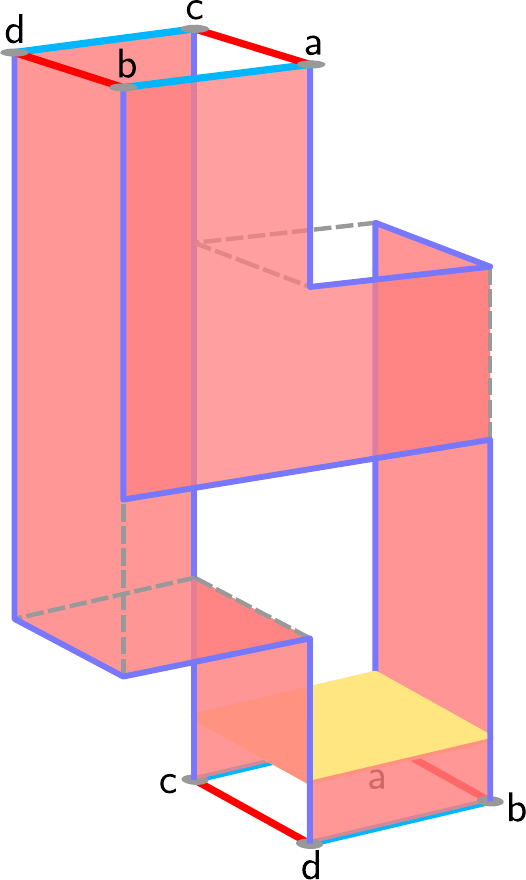}
        \caption{}
        \label{fig:Corner_mvmts}
    \end{subfigure}\hspace{70pt}
    \begin{subfigure}[b]{0.21\textwidth}
    \centering\includegraphics[width=\textwidth]{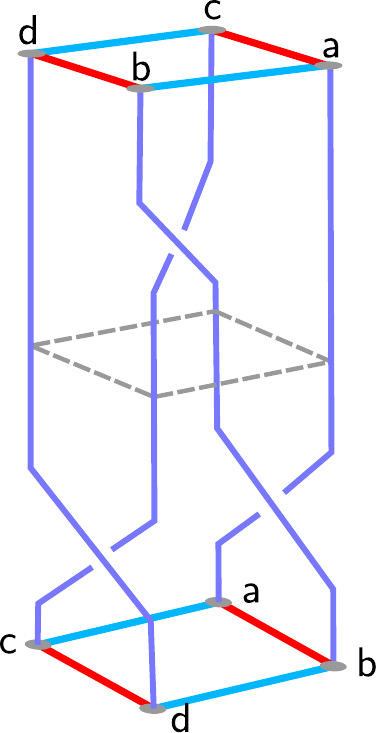}
    \caption{}
    \label{fig:braid_diagram}
    \end{subfigure}
    \caption{The procedures illustrated in \Cref{fig:ST_diagrams}, one of which is reproduced in (a), are shown to be topologically equivalent to a sequence of braids of the logical corners (blue wires), shown in (b).}
    \label{fig:space-time_diagrams}
\end{figure}

As can be seen from the schematic illustration of this procedure in \Cref{fig:twist_braid_schem}, the result of these measurements is the rotation of the logical corners by $90^\circ$ in the clockwise direction. We also display, in \Cref{fig:space-time_diagrams}, how the logical corner movements can be interpreted as a sequence of braids that result in the same $90^\circ$ rotation. In \Cref{fig:Corner_mvmts}, we reproduce one of the diagrams from \Cref{fig:ST_diagrams} for clarity. \Cref{fig:braid_diagram} shows the same procedure but with the movements rearranged into a series of clockwise exchanges. The domain wall is removed for clarity.

The sequence of braids thereby enacted is equivalent (up to a logical Pauli operator) to the braid transformation introduced in Reference~\cite{Brown_2017} that produces a logical Hadamard on the logical qubit. Call $B_1$ the clockwise braid of the bottom pair of corners (c and d in the first panel of \Cref{fig:twist_braid_schem}), $B_2$ the clockwise braid of the top pair of corners (a and b) and $B_3$ the clockwise braid of diagonally opposite corners (a and d). Then, both procedures from \Cref{sec:explanatation} can be seen to enact the braid $B_3 B_2 B_1$, as shown in \Cref{fig:braid_diagram}. Note that $B_2 B_1$ is equivalent to a logical Pauli operator. One can see this by noting that $B_2$ and $B_1$ are logically equivalent to one another. Then $B_2 B_1 \equiv B_1^2 \propto \bar{X}$ (completing a braid twice is equivalent to the logical operator encircling the two braided twists, owing to their similarity to Ising anyons~\cite{Bombin_2010}). By tracking logical operators, one can see that $B_3$ is equivalent to $\bar{H}$ up to some logical Pauli found by keeping track of the signs of the logical operators before and after the braid. Therefore, the combination of braids shown in \Cref{fig:braid_diagram} can be seen to, in fact, enact the logical Hadamard gate (up to a logical Pauli gate, but this is, in the present case, just the identity gate in both procedures).

\section{Hook errors in patch extension steps mean that $d$ QEC rounds are required}\label{app:hook_error}

\begin{figure}
    \centering
    \begin{subfigure}[b]{0.3\textwidth}
        \includegraphics[width=\textwidth]{LHDW_patch_g.pdf}
        \caption{\label{fig:LHDW_g_reprod}}
    \end{subfigure}\hfill
    \begin{subfigure}[b]{0.25\textwidth}
        \includegraphics[width=\textwidth]{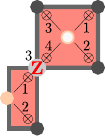}
        \caption{\label{fig:Hook_error_schedule}}
    \end{subfigure}\hfill
    \begin{subfigure}[b]{0.23\textwidth}
        \includegraphics[width=\textwidth]{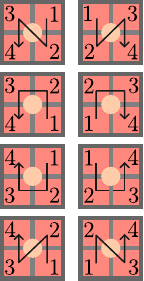}
        \caption{\label{fig:Possible_X_schedules}}
    \end{subfigure}\\
    \begin{subfigure}[b]{0.4\textwidth}
    \includegraphics[width=\textwidth]{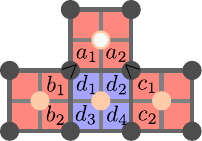}
    \caption{\label{fig:app_schedules_around_Z_plaq}}
    \end{subfigure}\hfill
    \begin{subfigure}[b]{0.4\textwidth}
        \includegraphics[width=\textwidth]{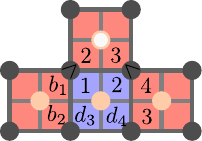}
        \caption{\label{fig:app_schedules_around_Z_plaq_2}}
    \end{subfigure}
    \caption{(a) Reproduction of \Cref{fig:LHDW_g} for convenience. (b) An example of a hook error that can reduce the effective distance from $k+1$ to $k$ after $k$ QEC rounds with the patch from (a). $CX$ gates are numbered according to the layer in which they occur. The $Z$ error occurs on the data qubit indicated during the 3\textsuperscript{rd} layer (in the first QEC round). (c) The possible schedules for $X$ stabilisers are shown. Numbers in the quadrants indicate the order of the elements in the permutation shown by the arrow. (d) A small part of the patch in (a), involving a non-deterministic stabiliser, is shown. The main text shows that, under standard assumptions, $a_1 < b_1$ and $a_2<c_1$ produces a contradiction. (e) One possible scheduling that violates the standard scheduling assumptions.} 
    \label{fig:app_B}
\end{figure}

We explained in \Cref{sec:distance_repeat_mmts} that the first patch extension step (\Cref{fig:LHTrans_c} and \ref{fig:LHDW_c}) and the corner movement step (\Cref{fig:LHTrans_d} 
and \ref{fig:LHDW_d}) both require $d$ QEC rounds under phenomenological noise, which also suffices under circuit-level noise. However, we pointed out that, in the last patch extension step (\Cref{fig:LHDW_g}, also reproduced in \Cref{fig:LHDW_g_reprod} for convenience), only $d-1$ QEC rounds are needed under phenomenological noise to maintain the distance of the logical Hadamard procedure. However, it turns out that, under circuit level noise, we still need $d$ QEC rounds for the last patch extension step to maintain effective distance of the logical Hadamard procedure, given certain assumptions on the scheduling. This is due to hook errors that have both space and time components. We explain the details of this below, and note that our observations were also verified using \texttt{stim} \cite{stim}.

We first explain that, with $d-1$ QEC rounds under circuit-level noise, only $d-1$ faults are required to produce an undetected logical error in the patch extension step of \Cref{fig:LHDW_g_reprod}. The $X$ stabilisers are measured using an ``N''-shaped schedule during this stage of the procedure (see \Cref{sec:Stab_schedules}). Hence, the qubit at $(1,9)$ is hit by entangling gates from $X$ stabiliser circuits during the first layer and the fourth layer of the first QEC round, as shown in \Cref{fig:Hook_error_schedule}. Therefore, if a $Z$ error occurs on the data qubit between these two layers (say, the third), it will only flip the $(0,8)$ stabiliser in the second (and subsequent) QEC rounds. Meanwhile, it will flip both the green logical observable and all measured values of the $(2,10)$ stabiliser. Therefore, it will not trigger any detectors at the location $(2,10)$, since the original measured value of the stabiliser was non-deterministic. This error is also shown in \Cref{fig:Hook_error_schedule}, happening in the third layer. If we only perform $d-1$ QEC rounds and this $Z$ error is followed by $d-2$ measurement errors of the $(0,8)$ stabiliser (a total $d-1$ number of faults), no detector is triggered and hence the flip of the green logical observable is not detected. Hence, maintaining the effective distance requires $d$ rounds of error correction during this patch extension step as well.

We now show that no alternative schedule (uniform across QEC rounds) with only four entangling layers will reduce the number of required QEC rounds back to $d-1$ under circuit-level noise without increasing the idling time of the qubits or sacrificing the effective distance of the patch in some other way. Let $S_\text{non-det}$ be the set of non-deterministic $X$ stabilisers and $S_\text{det}$ the set of deterministic $X$ stabilisers in the first QEC round in this patch extension step (those with/without white auxiliary qubits in \Cref{fig:LHDW_g_reprod} respectively). Consider qubit $q$ which is hit by one entangling gate from the auxiliary circuit of an $S_\text{non-det}$ stabiliser (call it $C_{q,\text{non-det}}$) and one from that of an $S_\text{det}$ stabiliser (call it $C_{q,\text{det}}$). A distance-reducing hook error (meaning that $d$ rounds are required) is possible whenever $C_{q,\text{det}}$ occurs \emph{before} $C_{q,\text{non-det}}$ for some qubit $q$, as described above (see \Cref{fig:Hook_error_schedule}). In that case, a $Z$ error occurring on $q$ between the two entangling gates will flip the non-deterministic stabiliser but not the deterministic one in that round. Thus we would like to find a schedule where all such qubits $q$ are hit by $C_{q,\text{non-det}}$ before $C_{q,\text{det}}$. In addition, we require $X$-type stabilisers to follow schedules that spread $X$ errors vertically, so that the effective $X$ distance of the patch is $d$. Consequently, the $X$ stabilisers have to follow one of the schedules from \Cref{fig:Possible_X_schedules}. Here, we assume that the scheduling is the same in each QEC round; without this assumption, alternative schedules apart from those in \Cref{fig:Possible_X_schedules} can be used without dramatically sacrificing the distance. However, while we do not prove our claim for this broader case, it remains unlikely that one can achieve distance $d$ for the logical Hadamard procedure with only $d-1$ QEC rounds in the last patch-extension step. 

As for $Z$-type stabiliser schedules, we only require that they are compatible with $X$-type schedules, namely that they satisfy the standard scheduling rules presented in \cite[Section 2]{tangled_schedules} and \cite[Figure 15]{LitOpp}. We comment more on these scheduling rules below. We will see that these requirements are incompatible if we assume four layers of entangling gates (that is, we do not include any additional idling layers).

Consider the stabilisers shown in \Cref{fig:app_schedules_around_Z_plaq}: we have one non-deterministic stabiliser (which we will label $a$), two other deterministic $X$ stabilisers ($b$ and $c$) and one $Z$ stabiliser ($d$). Each stabiliser is drawn with four quadrants, each carrying a number (some of which are not included for clarity) indicating the layer of that stabiliser's auxiliary circuit in which the entangling gate at that location is performed. To avoid the possibility of distance-reducing hook errors, we require $b_1 > a_1$ and $c_1 > a_2$, as shown in the figure, and we also require that elements of vertical columns in $X$ stabilisers belong to one and only one of either $\lbrace 3,4\rbrace$ or $\lbrace 1,2\rbrace$ (see \Cref{fig:Possible_X_schedules}). For example, $b_1$ and $b_2$ must either be $1$ and $2$, or $3$ and $4$ in some order, so that the hook error of the type shown in \Cref{fig:error_spreading} at worst affects vertically adjacent qubits and hence the effective distance is not reduced. In order for the scheduling to be valid, we further require that, for example, $(d_1 < a_1) \land (d_2 < a_2)$ or $(d_1 > a_1) \land (d_2 > a_2)$, and similarly for other adjacent $Z$/$X$ stabilisers \cite{tangled_schedules,LitOpp}. Finally, if we assume no idling (that is, the number of two-qubit gate layers is four) and that no qubit can be acted on by two entangling gates in the same layer, we require all quadrants of a plaquette to be filled with unique numbers between $1$ and $4$ and all quadrants around a data qubit to also be filled with unique numbers in this range.

Now, we claim that we hereby produce a contradiction. Note the possibilities for $a_1$ and $a_2$. Since neither of them can be $4$ (to allow both $b_1 > a_1$ and $c_1>a_2$) and $a_1\in \lbrace 1,2\rbrace $ if and only if $a_2 \in \lbrace 3,4\rbrace$, the only possibilities are $(a_1,a_2) \in \lbrace (1,3),(3,1),(2,3),(3,2)\rbrace$. Since the setup is unchanged when we perform a reflection along the vertical axis of symmetry (under which $a_1 \leftrightarrow a_2$), we have only two distinct possibilities. As the first case, suppose $a_1 = 1$, $a_2 = 3$ holds. Then $c_1 = 4$ is forced, and so $c_2 = 3$ (from the possible $X$ schedules). Therefore, $d_2$ cannot be $3$ or $4$ and so we must have that $d_2 < a_2$ and $d_1 < a_1$, by the scheduling rules. But this is impossible, since $a_1 = 1$.

The other distinct possibility is $(a_1,a_2) = (2,3)$. The argument here proceeds in a similar way to produce a contradiction. In this case, one can see that the schedule shown in \Cref{fig:app_schedules_around_Z_plaq_2} is the only one consistent with $b_1 > a_1$, $c_1>a_2$, and the scheduling rules, where $b_1,b_2 \in \lbrace 3, 4\rbrace$. If we choose $b_1 = 4, b_2 = 3$, then $d_3 = 4, d_4 = 3$ is forced, which is impossible because $c_2 = 3$. Meanwhile, $b_1 = 3, b_2 = 4$ forces $d_3 = 3$, $d_4 = 4$, which results in improper scheduling between stabilisers $d$ and $c$. We therefore arrive at a contradiction.

\end{document}